\documentclass[12pt,english]{article}
\pdfoutput=1
\usepackage{lmodern}

\usepackage[T1]{fontenc}
\usepackage[utf8]{inputenc}
\usepackage{geometry}
\usepackage{color}
\usepackage{babel}
\usepackage{mathtools}
\usepackage{amsmath}
\usepackage{ulem}
\usepackage{amssymb}
\usepackage{graphicx}
\usepackage{epstopdf}
\usepackage{esint}
\usepackage{subfigure}
\usepackage{setspace}
\usepackage[unicode=true,pdfusetitle,
 bookmarks=true,bookmarksnumbered=false,bookmarksopen=false,
 breaklinks=false,pdfborder={0 0 1},backref=false,colorlinks=true]
 {hyperref}
\hypersetup{
 citecolor=black,linkcolor=black,urlcolor=black}
\makeatletter

 \@ifundefined{textcolor}{}
 {
   \definecolor{BLACK}{gray}{0}
   \definecolor{WHITE}{gray}{1}
   \definecolor{RED}{rgb}{1,0,0}
   \definecolor{GREEN}{rgb}{0,1,0}
   \definecolor{BLUE}{rgb}{0,0,1}
   \definecolor{CYAN}{cmyk}{1,0,0,0}
   \definecolor{MAGENTA}{cmyk}{0,1,0,0}
   \definecolor{YELLOW}{cmyk}{0,0,1,0}
 }
\usepackage{babel}
\makeatletter
\def\simgt{\mathrel{\lower2.5pt\vbox{\lineskip=0pt\baselineskip=0pt
           \hbox{$>$}\hbox{$\sim$}}}}
\def\simlt{\mathrel{\lower2.5pt\vbox{\lineskip=0pt\baselineskip=0pt
           \hbox{$<$}\hbox{$\sim$}}}}
\makeatother

\newcommand{\bea}{\begin{eqnarray}}
\newcommand{\eea}{\end{eqnarray}}
\newcommand{\eq}[1]{\begin{align}#1\end{align}}
\newcommand{\Ref}[1]{Ref.~\cite{#1}}
\newcommand{\Refs}[1]{Refs.~\cite{#1}}
\newcommand{\Fig}[1]{Fig.~\ref{#1}}
\newcommand{\Eq}[1]{Eq.~\eqref{#1}}
\newcommand{\Eqs}[2]{Eqs.~\eqref{#1} and \eqref{#2}}
\newcommand{\Sec}[1]{Sec.~\ref{#1}}

\newcommand{\App}[1]{App.~\ref{#1}}

\newcommand{\Veff}{V_{\rm eff}}
\newcommand{\mPl}{m_{\rm Pl}}

\newcommand{\OO}{\mathcal{O}}

\newcommand{\eV}{\text{ eV}}

\newcommand{\GeV}{\text{ GeV}}
\newcommand{\TeV}{\text{ TeV}}

\newcommand{\Vv}{V_{\rm vac}}
\newcommand{\Vvc}{V_{\rm vac,0}}
\newcommand{\Vvq}{\Delta V_{\rm vac}}
\newcommand{\Vmod}{V_{\rm mod}}
\newcommand{\VH}{V_H}

\newcommand{\vobs}{v_{\rm obs}}
\newcommand{\vmax}{v_{\rm max}}
\newcommand{\Hinf}{H_{\rm inf}}
\newcommand{\Vinf}{V_{\rm inf}}
\newcommand{\taudiff}{\tau_{\Delta H}}
\newcommand{\Ne}{\mathcal{N}_{\rm e-folds}}
\newcommand{\Nobs}{\mathcal{N}_{\rm obs}}
\newcommand{\lambdaH}{\lambda}
\newcommand{\eg}{{\it e.g.}}

\begin{document}
	\interfootnotelinepenalty=10000
	\baselineskip=18pt
	\hfill CALT-TH-2018-048
	\hfill
	
	\vspace{2cm}
	\thispagestyle{empty}
	\begin{center}
		{\LARGE  \bf
			 Mass Hierarchy and Vacuum Energy
		}\\
		\bigskip\vspace{1.cm}{
			{\large Clifford Cheung${}^a$ and Prashant Saraswat${}^b$}
		} \\[7mm]
		{\it Walter Burke Institute for Theoretical Physics, \\[-1mm]
			California Institute of Technology, Pasadena, CA 91125
			}\let\thefootnote\relax\footnote{${}^a$e-mail: \url{clifford.cheung@caltech.edu} \\
			\indent \hspace{2mm} ${}^b$e-mail: \url{saraswat@caltech.edu} } \\
	\end{center}
	\bigskip
	\centerline{\large\bf Abstract}

\begin{quote} \small
 
 A hierarchically small weak scale does not generally coincide with enhanced symmetry, but it may still be exceptional with respect to vacuum energy.   By analyzing the classical vacuum energy as a function of parameters such as the Higgs mass, we show how near-criticality, {\it i.e.}~fine-tuning, corresponds universally to boundaries where the vacuum energy transitions from exactly flat to concave down.  In the presence of quantum corrections, these boundary regions can easily be perturbed to become maxima of the vacuum energy. After introducing a dynamical scalar field $\phi$ which scans the Higgs sector parameters, we propose several possible mechanisms by which this field could be localized to the maximum. One possibility is that the $\phi$ potential has many vacua, with those near the maximum vacuum energy expanding faster during a long period of cosmic inflation and hence dominating the volume of the Universe. Alternately, we describe scenarios in which vacua near the maximum could be anthropically favored, due to selection of the late-time cosmological constant or dark matter density. Independent of these specific approaches, the physical value of the weak scale in our proposal is generated naturally and dynamically from loops of heavy states coupled to the Higgs. These states are predicted to be a loop factor heavier than in models without this mechanism, avoiding tension with experimental null results. 
\end{quote}

\setcounter{footnote}{0}

\newpage
\tableofcontents
	 
\newpage

\section{Introduction}
\label{sec:intro}

The colossal hierarchy between the weak scale and the Planck scale has been a central inspiration for physics beyond the standard model (SM) for many decades.  Considerable effort has been spent building models that temper the electroweak hierarchy with additional symmetries of nature which invariably entail new light degrees of freedom.  Unfortunately, the long march of null results from the Large Hadron Collider (LHC) has cast doubt on the viability of these models, and in turn the original motivations of naturalness.

At such a remarkable juncture there is an impetus to revisit naturalness as a physical criterion. Unfortunately, the concept of naturalness is  often illustrated with a cartoonish picture in which a fundamental scalar is light because formally infinite quantum corrections to its mass are delicately balanced against bare mass counterterms in the Lagrangian.  This picture is lacking as both infinite quantities are scheme dependent and neither is actually observable, even in principle.  Only their sum, the physical mass, has invariant meaning.  Furthermore, by the usual rules of quantum field theory, when a counterterm for the mass exists, there is no recourse but to extract its value from some physical quantity.  In principle, this quantity might correspond to a boundary condition in the deep infrared {\it e.g.} such as a pole mass, or alternatively, a boundary condition in the ultraviolet such as a model parameter at short distances.  The latter choice is reasonable if one assumes there is a completely reductive and unified theory of the ultraviolet in which the scalar mass is  {\it finite} and  {\it calculable} from independent theory parameters.  In this case the scalar mass is a {\it prediction} deduced from other observables and need not be measured independently.  In such a context there is an honest notion of unnaturalness, \eg, for a scalar whose calculable mass is tiny despite strong sensitivity to the theory parameters defined in the ultraviolet.  We adopt this viewpoint here.

The intractability of the hierarchy problem is tightly linked to the fact that the near-criticality of a small weak scale does not generally correspond to a point of enhanced symmetry. In this paper, we argue that while a small weak scale is not special with respect to symmetry, it {\it can} be special with respect to energy in a broad class of models.  Our logic is analogous to the axion solution of the strong CP problem.  As shown by Vafa and Witten~\cite{Vafa:1984xg}, the vacuum energy of QCD is minimized for vanishing $\theta$ parameter, indicating that the ground state of the theory is parity conserving.  When $\theta$ is promoted to the QCD axion field, the theory automatically relaxes to this minimum.  From this perspective, the success of the QCD axion hinges critically on the fact that the vacuum energy of QCD is minimized for vanishing $\theta$.

Here we ask: what is the analogous behavior of the vacuum energy as the electroweak scale is varied? To be precise, consider the SM, where the vacuum energy receives contributions both from electroweak symmetry breaking (EWSB) as well as quantum corrections.   In a fixed model the vacuum energy is of course an absolute constant, but we can think of it as a function of other ``fundamental'' parameters in the theory.  If we introduce a dynamical field $\phi$ whose zero-mode value modulates these parameters, then the vacuum energy simply becomes the effective potential for $\phi$.  For example in the SM, $\phi$ might control the Higgs mass by coupling as $\phi |H|^2$. Integrating out all the SM particles generates a potential, $\Vv(\phi)$, describing how the vacuum energy varies with respect to the parameter that $\phi$ scans.  As we will show, at the classical level the vacuum energy has a {\it universal} structure as a function of $\phi$: a plateau along directions of unbroken electroweak symmetry, bounded by decreasing, concave down regions where symmetry breaking occurs.  Hence, near-criticality always occurs when $\phi$ approaches the ``cliff's edge'' of the classical vacuum energy.

Of course, the classical vacuum energy is corrected by quantum effects which are far from negligible and are in fact the new guise of the hierarchy problem in this context. Indeed, the radiative instability of $\Vv(\phi)$ reflects the fact that vacuum energy is sensitive to ultraviolet scales as well as low-energy parameters scanned by $\phi$. For example, any heavy particles which couple to the Higgs will induce one-loop corrections to its mass. If $\phi$ only couples to heavy states through loops of the Higgs, the vacuum energy  $\Vv(\phi)$ is first corrected at two loops.\footnote{It is crucial for our setup that the dominant quantum corrections to $\Vv(\phi)$ are mediated by the Higgs.  This occurs if $\phi$ has a shift symmetry broken only by its coupling to the Higgs.  In this sense the Higgs acts as a ``messenger'' communicating ultraviolet mass scales to $\phi$.  
This assumption is violated if $\phi$ has a sizable bare potential to start or couples directly to heavy states.  Throughout, we take a conservative albeit legalistic approach of technical naturalness whereby the bare potential for $\phi$ is taken to be of order the leading radiative corrections which generate it.   }

Consequently, the universal classical structure described is robust provided that all states that couple directly to the Higgs are heavier by at most two loops in mass squared, corresponding to $\Lambda \lesssim$ 10 TeV.    As we will see, when the heavy particles roughly saturate this bound, their radiative corrections generically tilt the classical potential so that the observed value of the weak scale corresponds to the {\it absolute maximum} of the full quantum-corrected vacuum energy.  For obvious reasons, we will henceforth refer to this region as the ``mountaintop''.

The fact that a small weak scale can coincide with the maximum of the vacuum energy suggests a possible underlying ``principle of maximal vacuum energy" in nature.  But what reason have we to live on the mountaintop?  After all, fields automatically roll away from such maxima.  To remedy this, we add to the theory a bare potential $\Vmod(\phi)$ with many small bumps across field space, {\it e.g.}~as would arise from a periodic potential.  In general this will produce local vacua near the mountaintop.  Still, the question remains of how our present day Universe came to reside at such a place.  To arrive at the summit, we discuss several alternative mechanisms, ranging from conservative to exotic:
\begin{itemize}
\item
In the simplest approach, we may take the height of $\Vmod(\phi)$ to be sufficiently small, such that there exist local minima only very close to the ``summit'' of the mountaintop potential $\Vv$, where the weak scale is low, and not on the ``slopes'' with larger Higgs mass. Then one may argue that the weak scale takes on the observed value because there exist no other nearby vacua with a different value. This argument does not however explain why the Universe realizes the initial condition placing $\phi$ near a stable vacuum on the mountaintop, instead of rolling away from it to some distant value.

\item In order to explain why the mountaintop is a preferred initial condition for the observable universe,  we can utilize a recent and interesting proposal of Geller, Hochberg and Kuflik \cite{Geller:2018xvz}, who attack the hierarchy problem by exploiting the fact that regions of higher vacuum energy expand faster during inflation and thus populate greater volumes within the Universe. Since the maximum of $\Vv(\phi)$ corresponds to the observed weak scale in our approach as well, this inflationary attractor mechanism can be implemented.
  
\item Another option is to use anthropic reasoning to motivate residing on the mountaintop.  For example, if $\phi$ controls the late-time dark matter density, then rolling off of the mountaintop could constitute a catastrophic boundary.  Another possibility is that the cosmological constant is scanned by some additional landscape of degrees of freedom, {\it e.g.}~four-form fluxes, which are exponentially peaked at large values.  In this case the anthropic constraint on the cosmological constant makes it overwhelmingly likely that we reside at the maximum of $\Vv(\phi)$.

\item Last of all, one could imagine dynamically settling to the mountaintop via more exotic mechanisms.  If $\phi$ is promoted to a ghost field then it will roll simply roll to the maximum.   Another possibility is to introduce certain non-local operators involving the Higgs.  While consistent, this operator approach is peculiar since there are no dynamical modes involved.

\end{itemize}
The mechanisms for localizing our Universe to the maximum of $\Vv(\phi)$ are quite different but they all share a common experimentally verifiable prediction.  As noted earlier, the heavy states of mass $\sim$ few TeV generate radiative corrections that tilt the classical potential in a way that {\it defines} the exact value of the weak scale.  In this sense there is a tight correlation between the properties of the new physics and the low-energy parameters.

\section{Criticality and Vacuum Energy}
\label{sec:crit}

In this section we discuss the generic behavior of the vacuum energy $\Vv(\phi)$ as $\phi$ scans the theory through a critical point. We will decompose the vacuum energy into two components, a piece $\Vvc$ arising at the classical level and a piece $\Vvq$ generated by quantum corrections:
\eq{
\Vv(\phi) = \Vvc(\phi) + \Vvq(\phi)
\label{eq:Vvdef}
}
We will see that the classical vacuum energy $\Vvc$ has universal behavior: it is is flat in regions where symmetry is restored and slopes downward whenever symmetry breaking occurs. The quantum effects $\Vvq$ typically alter this, but in a way that generically turns the critical boundary between regions into maxima of the total vacuum energy $\Vv(\phi)$.

\subsection{Classical Potential} \label{sec:classical}

To begin, we work at the level of the classical potential, where the vacuum energy is dictated purely by the equations of motion of dynamical fields.

\subsubsection{General Argument}
  For simplicity, we study a theory of a real scalar $H$ which depending on the potential, may or may not acquire a vacuum expectation value (VEV) and induce spontaneous symmetry breaking.   Denoting the $H$ potential by $\VH(H)$, we also add a scanning field $\phi$ which modulates some $H$ dependent operator $\OO(H)$ such that $\OO(0)=0$.  This guarantees that $\OO(H)$ is an order parameter for the VEV of $H$.  The full potential for $H$ and $\phi$ is
\eq{
V(H, \phi) = \VH(H) + \phi \,  {\cal O}(H),
\label{eq:scan}
}
For example, taking $\OO(H) \propto H^2$ corresponds to scanning the Higgs mass, although as we will discuss later, it is possible to scan any parameter in the potential.

Let us define $v(\phi) \equiv  \langle H(\phi) \rangle$ to be the VEV of $H$ at fixed value of $\phi$, determined by solving the zero mode equation of motion for $H$,
\eq{
\VH'(v(\phi)) + \phi  \, {\cal O}'(v(\phi)) =0
\label{eq:eom}
} 
We define $\Vvc(\phi)$ as the classical vacuum energy as a function of $\phi$ after integrating out $H$.  Here it suffices to plug the VEV of $H$ back into the original potential,
\eq{
\Vvc(\phi) = V(v(\phi),\phi) =  \VH(v(\phi))  + \phi \,   {\cal O}(v(\phi)) .
}

We can ascertain the critical points of $\Vvc$ by taking its derivative with respect to $\phi$
\begin{align}
\Vvc'(\phi)  =\OO(v(\phi)).
\label{eq:Vp}
\end{align}
where we have used the $H$ equation of motion in \Eq{eq:eom}. In other words, the slope of the vacuum energy is exactly the VEV of the operator ${\cal O}$. Consequently, $\Vvc$ slopes for values of $\phi$ in which symmetry is broken and is flat for values of $\phi$ in which symmetry is restored.

Taking a second derivative of the vacuum energy, we obtain. 
\eq{
\Vvc''(\phi) = {\cal O}'( v(\phi))  v'(\phi) 
}
Compare this to the first derivative with respect to $\phi$ of the $H$ equation of motion in \Eq{eq:eom}, 
\eq{
\left[ \VH''(v(\phi)) + \phi  \,  \OO''(v(\phi)) \right]  v'(\phi)  +  \OO'(v(\phi))= 0.
}
Here we observe that the quantity in the square brackets is simply $m^2(\phi)$, the squared mass of the physical $H$ fluctuations about the vacuum.  Solving for $\OO'(v(\phi))$ and plugging back into our expression for $\Vvc(\phi)''$ yields
\eq{
\Vvc''(\phi) = -  m^2(\phi)   v'(\phi)^2 \leq 0. 
\label{eq:Vpp}
} 
The inequality follows because the $H$ fluctuations about the vacuum should have positive mass for a stable minimum. 

\begin{figure}
\begin{center}
  \includegraphics[width=17cm]{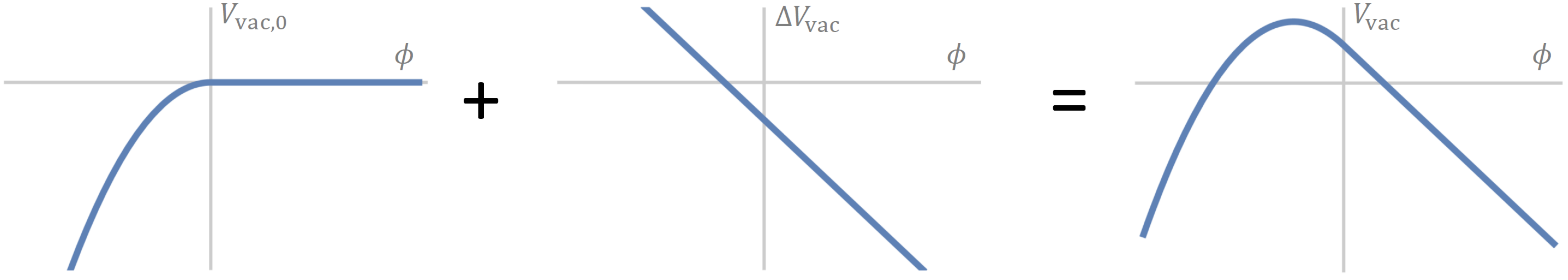}
 \end{center}
 \caption{Illustrative sketch of the contributions to the effective potential as a function of a field $\phi$ which linearly scans the Higgs potential. The leftmost plot shows the classical potential $\Vvc$ for $\phi$ discussed in \Sec{sec:classical}, which is zero for unbroken symmetry ($m^2(\phi) > 0$) and concave down in the broken symmetry region ($m^2(\phi) < 0$). The middle plot shows the quantum potential $\Vvq$ (\Sec{sec:quantum}) induced for $\phi$ (for one possible sign), which is featureless in the neighborhood of $m^2(\phi) = 0$. The rightmost plot shows the combination $\Vv(\phi)$, which gives a maximum for the vacuum energy at a parametrically low value of $m^2(\phi)$ and $\langle H \rangle$.
}
 \label{fig:plateau}
 \end{figure}

The above results imply a universal shape for the classical vacuum energy $\Vvc(\phi)$ when $\phi$ linearly couples to some order parameter. It is completely flat for values of $\phi$ for which the symmetry is unbroken.  As shown earlier, the slope of the vacuum energy is equal to VEV of the order parameter to which $\phi$ couples.  For values of $\phi$ at which the symmetry is broken, the vacuum energy slopes downwards with negative curvature. The result is sketched in the leftmost plot of \Fig{fig:plateau}.

The potential defined in \Eq{eq:scan} naively requires a quite specific choice of linear coupling of $\phi$ to a single operator $\OO$.  However in the context of smoothly varying scalar fields in effective field theory, this choice is strongly motivated. In particular, consider that the field $\phi$ can a priori couple via a more complicated interaction $\sum_i f_i(\phi) {\cal O}_i(H)$ for various functions $f_i$ and operators ${\cal O}_i$ which are assumed to be order parameters for the phase transition. 

We can restrict our attention to the lowest dimension operator coupling $\phi$ to an operator $\OO_i$, which is a linear coupling.  Equivalently, we can Taylor expand the function $f_i(\phi)$ to linear order around the critical value $\phi_\text{crit}$ where symmetry breaking occurs. The above assumption is in fact exactly those of the Landau theory of phase transitions, used to describe universal behavior near a critical point. Therefore in at least the Landau theory limit, some vicinity of the symmetry breaking point, the form of Eq.~\ref{eq:scan} applies, as do the results of Eqs.~\ref{eq:Vp}~and~\ref{eq:Vpp}. 

\subsubsection{Standard Model}
 
 As an example of the above, let us take the SM Higgs potential $\VH(H) = \mu_0^2 |H|^2 + \lambda_0 |H|^4$ and introduce a scalar field $\phi$ which couples to the operator ${\cal O}(H) = g |H|^2 + \kappa |H|^4$ for some coupling constants $g$ and $\kappa$.  We have shifted notation to have $H$ represent the usual complex $SU(2)$ doublet Higgs. Let us take $v(\phi)  \equiv  \sqrt{2 H^\dagger H}$; i.e. $v(\phi)$ is the vev of a real scalar component of $H$. The $\phi$ field simultaneously scans the Higgs mass and quartic coupling linearly. The potential can be written as
\eq{
V(H,\phi) =  \mu^2(\phi) |H|^2 + \lambdaH(\phi) |H|^4, \label{eq:tachyon}
}
where $\mu^2(\phi) = \mu_0^2 + g \phi$ and $\lambdaH(\phi) = \lambda_0 + \kappa \phi$. Let us assume $\lambdaH(\phi) > 0$ for all $\phi$ of interest so that the potential is bounded. The Higgs VEV is then expressed as
\eq{
v(\phi)= 
\left\{ \begin{array}{ll}
0  &,\quad \mu^2(\phi) \geq 0 \\
\sqrt{- \dfrac{\mu^2(\phi)}{\lambdaH(\phi)}} &,\quad \mu^2(\phi)<0
\end{array}
\right. 
}
Plugging back into the potential, we obtain the vacuum energy,  
\eq{
\Vvc(\phi) = 
\left\{ \begin{array}{ll}
0  &,\quad \mu^2(\phi) \geq 0 \\
-\dfrac{\mu^4(\phi)}{4\lambdaH(\phi)} &,\quad \mu^2(\phi)<0
\end{array}
\right. .
\label{eq:VeffEWSB}
}
The first derivative of the vacuum energy in the EWSB phase is 
\eq{
\Vvc'(\phi)   = \frac{1}{2} g v(\phi)^2 + \frac{1}{4} \kappa v(\phi)^4 =\OO(v(\phi)) 
}
in accordance with \Eq{eq:Vp}. Furthermore the second derivative is 
\eq{
\Vvc''(\phi) = -\frac{1}{2\lambdaH (\phi)} \left[ g - \frac{\kappa \mu^2(\phi)}{\lambdaH(\phi)}  \right]^2< 0
}
so that for any value of $g$ and $\kappa$ the potential is concave down. 
 
\subsection{Quantum Potential} \label{sec:quantum}

Thus far we have completely neglected the crucial effects of quantum corrections, which are of course the entire origin of the hierarchy problem. The potential in \Eq{eq:scan} which we have assumed thus  far is not stable under quantum corrections induced by short-distance fluctuations. In particular, we assumed that the field $\phi$ coupled only to an order parameter ${\cal O}(H)$ satisfying ${\cal O}(0) = 0$.  However, loops of $H$ particles will generate new operators which are independent of $H$ but dependent on $\phi$.  
In this section we consider these effects for the Higgs potential in \Eq{eq:tachyon} and compute the quantum corrected vacuum energy $\Vv$ as a function of $\phi$.

\subsubsection{Indirect Coupling}
 
In order to be very concrete about the physics of the sensitivity to ultraviolet scales, we will consider the case where the scalar $H$ couples to a new heavy particle of mass $\Lambda$ with $\OO(1)$ coupling constant. We do this to avoid irrelevant confusions having to do with ultraviolet regulators and scheme dependence. Independent of the regulator, the scalar $H$ receives a large, running contribution to its mass proportional to $\Lambda$, reflecting the hierarchy problem. Hence, we will take the physical threshold $\Lambda$ as an effective ultraviolet cutoff everywhere.

\begin{figure}
\begin{center}
\begin{tabular}{ccc}
   \includegraphics[width=5.5cm]{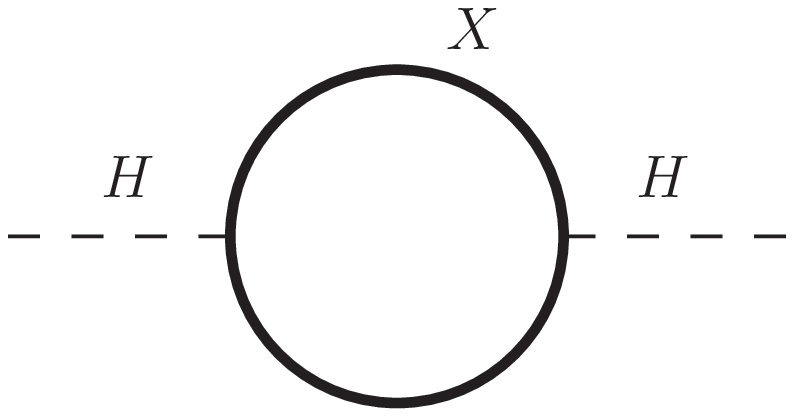}
   &
   \includegraphics[width=5.2cm]{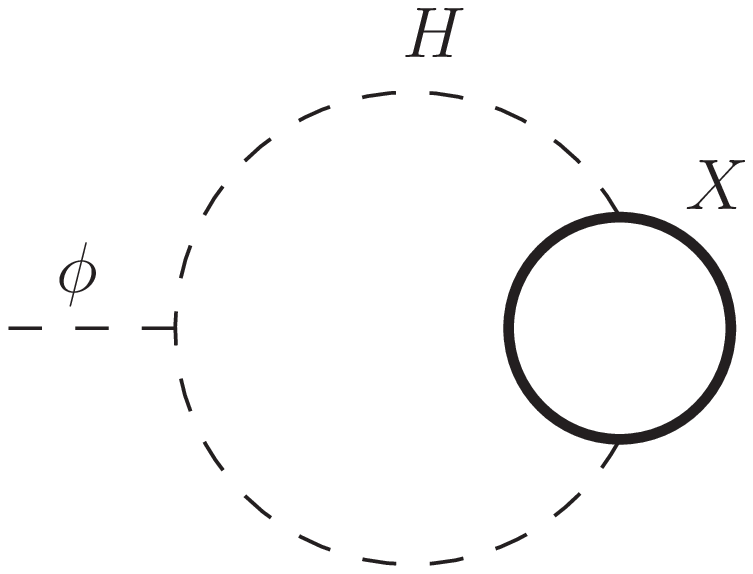}
   &
      \includegraphics[width=4cm]{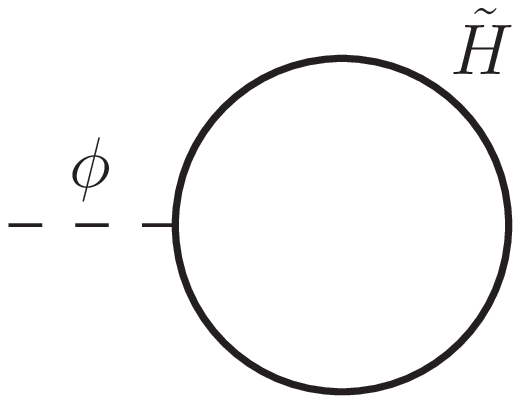}
 \\
 (a) 
 &
 (b) 
 &
 (c) 
 \end{tabular}
 \end{center}
 \caption{Feynman diagrams contributing to the Higgs and $\phi$ potentials. (a) One-loop Higgs mass correction due to Higgs couplings to a heavy particle $X$ (shown as a fermion here).  (b) Two-loop $\phi$ tadpole mediated through the Higgs from a heavy $X$ particle. (c) One-loop $\phi$ tadpole due to loops of a heavy field $\tilde{H}$, {\it e.g.} a Higgsino, that couples directly to $\phi$.
}
 \label{fig:feynman}
 \end{figure}

For now let us assume that the field $\phi$ does not couple at tree-level to the heavy states. Nevertheless, it will be sensitive to them through loops of $H$, as illustrated in \Fig{fig:feynman}. For simplicity let us consider $\kappa = 0$ in the model defined in \Eq{eq:tachyon}.  That is, $\phi$ scans only the Higgs mass via $\mu^2(\phi) = \mu_0^2 + g \phi$. Estimating the quantum corrected potential up to two loop order yields 
\eq{
\Delta V(H,\phi) =  - \Delta M^2 \mu^2(\phi)  +  \Delta \mu^2  |H|^2 +\ldots, \label{eq:Vq}
}
where we have defined the quantum corrections,
\eq{
\Delta M^2 \sim  \frac{ \Lambda^2}{(16\pi^2)^2}  \qquad \textrm{and} \qquad \Delta \mu^2  \sim \frac{\Lambda^2}{16\pi^2}, \label{eq:musq}
}
and the ellipses denote operators which we ignore because they are independent of $\phi$ or not power law divergent and thus not ultraviolet sensitive.   The above expressions hold up to $\OO(1)$ dimensionless coefficients which depend on the details of the ultraviolet completion. For reasons that will be later apparent, we will {\it assume} an ultraviolet completion in which $\Delta M^2$ is positive.  This binary choice does not require tuning.  

The quantum correction to the scalar mass, $\Delta \mu^2$, is generated by one-loop diagrams with a virtual heavy particle.  On the other hand, $\Delta M^2$, parameterizing the $\phi$ tadpole contribution to the vacuum energy term, is generated by two-loop diagrams with a virtual scalar and a virtual heavy particle. This term is an echo of the hierarchy problem of $H$ feeding in at higher loop into the bare potential for $\phi$.

Next, we integrate out $ H$, including both classical and quantum contributions.  Without loss of generality it will be convenient to redefine the classical mass \eq{
\mu^2(\phi) \rightarrow \mu^2(\phi) - \Delta \mu^2. \label{eq:mass_shift}
}
This absorbs the one-loop shift of the mass into the definition of $\mu^2(\phi)$, and shifts the $\phi$-independent vacuum energy by a constant.  The vacuum energy is then
\eq{
\Vvc(\phi) + \Vvq(\phi) =    - \Delta M^2 \mu^2(\phi)  +  \left\{ \begin{array}{ll}
0  &,\quad \mu^2(\phi) \geq 0 \\
-\dfrac{\mu^4(\phi)}{4\lambdaH} &,\quad \mu^2(\phi)<0
\end{array}
\right\}   + \ldots \label{eq:Veffq}
}
where the ellipses denote terms where are subdominant to those shown explicitly, either because they are independent of $\phi$ or because they arise from scalar loops and enter at ${\cal O}(\mu^4(\phi) /16\pi^2)$.

The quantum corrected potential has a maximum defined by the equation of motion,
\eq{ 
\Vv'(\phi_{\rm max})= \Vvc'(\phi_{\rm max}) +  \Vvq'(\phi_{\rm max}) =0,
}
where $\phi_{\rm max}$ is the value of $\phi$ at the maximum of the vacuum energy.  The solution is
\eq{
\mu^2( \phi_{\rm max}) = -2 \lambdaH \Delta M^2
}
The VEV of the Higgs ($v(\phi) = \sqrt{2} \langle H(\phi)\rangle$) at this maximum is then
\eq{
|v (\phi_{\rm max})|^2 = 2 \Delta M^2 \sim  \frac{ \Lambda^2}{\left(16\pi^2\right)^2}, \label{eq:phi_indirect}
 }
so the EWSB scale at this point is {\it determined} by the leading quantum correction to the bare $\phi$ potential.  Note that $\Delta M^2$ is two loop factors suppressed from from the heavy scale $\Lambda$, compared to the one-loop suppression of $\Delta m^2$. 

\subsubsection{Direct Coupling}

On the other hand, it is possible for the field $\phi$ to couple {\it directly} to new heavy particles.  Such a circumstance is motivated by supersymmetry (SUSY), where $H$ would naturally be embedded within a chiral superfield. In components, the Lagrangian would include terms like 

\eq{
-{\cal L} = (|\mu(\phi)|^2 - \Lambda^2) | H|^2 + \mu(\phi) \widetilde  H \widetilde  H + \textrm{c.c.},
}
 where $\mu(\phi) = \mu_0 + y \phi$ is a supersymmetric coupling that mediates scanning of the Higgs mass and we have lifted $ H$ into a chiral multiplet by introducing a fermionic superpartner $\widetilde  H$. Here $\Lambda$ parameterizes the mass splitting between states, and is thus a measure of SUSY breaking which also functions as an effective ultraviolet cutoff.   

At one loop, quadratically divergent corrections to the $\phi$ potential cancel by SUSY, while the finite and log-divergent terms are of order
\eq{ 
\Vvq =  - \Delta M^2 |\mu(\phi)|^2   + \ldots
}
where $\Delta M^2 \sim \Lambda^2/16\pi^2$.   If $\Delta M^2 > 0$, then $\Delta \Veff$ can again balance against the vacuum energy from EWSB to again create a maximum at
\eq{
|v(\phi_{\rm max})|^2 = 2 \Delta M^2 \sim  \frac{  \Lambda^2}{16\pi^2}, \label{eq:phi_direct}
}
which is now only one loop factor down from the heavy scale $\Lambda$.  Thus, when a direct coupling is present, then $\Lambda$ must be smaller in order to keep the weak scale light. 

\subsection{Resulting Mass Hierarchy}
\label{sec:hierarchy}

In \Sec{sec:classical} we computed in generality the vacuum energy $\Vv(\phi)$ produced by integrating out a field $H$ whose interaction parameters are effectively modulated by another scalar $\phi$.  As a function of $\phi$, the resulting vacuum energy $\Vv$ will generically have maxima near critical points where symmetry breaking occurs.  In \Sec{sec:quantum}, we found that the Higgs VEV at the maximum is $v^2 = \Delta M^2$, where depending on the model, $\Delta M^2$ is generated with either one-loop or two-loop suppression from some higher ultraviolet scale $\Lambda$. The observed small weak scale then places an upper bound on the scale $\Lambda$ and the masses of the associated states.     

For instance, consider the case of the SM with the Higgs mass modulated by $\phi$.  In the absence of additional symmetries, heavy particles which couple to the $H$ must have mass $ \lesssim 16 \pi^2 v_{\rm obs}$, where $v_{\rm obs}$ is observed EWSB scale, {\it i.e.}~$\sim 246 \GeV$ for the SM Higgs.  Moreover, the heavy particles which couple directly to $\phi$ must have mass $ \lesssim 4\pi v_{\rm obs}$. When these bounds are nearly saturated, the EWSB scale $v$ is actually dynamically generated by these heavy mass scales.

We can apply this order of magnitude logic to the minimal supersymmetric standard model (MSSM) as well.  If the SUSY mass term for the Higgs superfield is modulated by $\phi$, then we require the Higgsino and scalar Higgs partners to have mass $\lesssim 4\pi v_{\rm obs} \sim $ 1 TeV. This can be compared to the usual expectation that the Higgsino mass should be of order the weak scale $\sim 100 \GeV$ in the absence of tuning. Similarly the gauginos and sfermions can have masses up to $\sim 4 \pi$ larger than in the MSSM without fine-tuning, {\it e.g.}~the stops and gluinos could be as heavy as a few TeV. In \Sec{sec:SUSY} we will discuss a MSSM-like model and the resulting predictions in more detail.

\section{Minima on the Maximum}
\label{sec:minima}

We have studied a generic class of theories in which parameters in the Higgs sector are effectively modulated by a field $\phi$.  After integrating out the Higgs, we obtain a quantum corrected vacuum energy with a maximum corresponding to a parametrically small EWSB scale, $v_{\rm obs}$. For the SM model retrofitted with $\phi$, the vacuum energy becomes 

\eq{
\Vv(\phi) = 
\left\{ \begin{array}{ll}
- \lambdaH \left[v(\phi)^2 - \vobs^2 \right]^2/4 = -g^2 \phi^2/4\lambdaH  &,\quad \mu^2(\phi) \leq 0 \\[\medskipamount]
- g \vobs^2 \phi/2 + \lambdaH \vobs^4/4 &,\quad \mu^2(\phi)>0
\end{array}
\right. .
\label{eq:VvSM}
}
where we have chosen to define the location of the maximum, where $v(\phi) = \vobs$, as the origin for $\phi$. If one has a reason why $\phi$ should reside near the peak or ``mountaintop'' of this vacuum energy, then this would explain the smallness of the observed weak scale. A minimal condition for this is that $\phi$ is somehow stabilized near this region. In this section we present a simple model where vacua reside on the mountaintop.

\subsection{Model Definition}

\begin{figure}
\begin{center}
  \includegraphics[width=17.8cm]{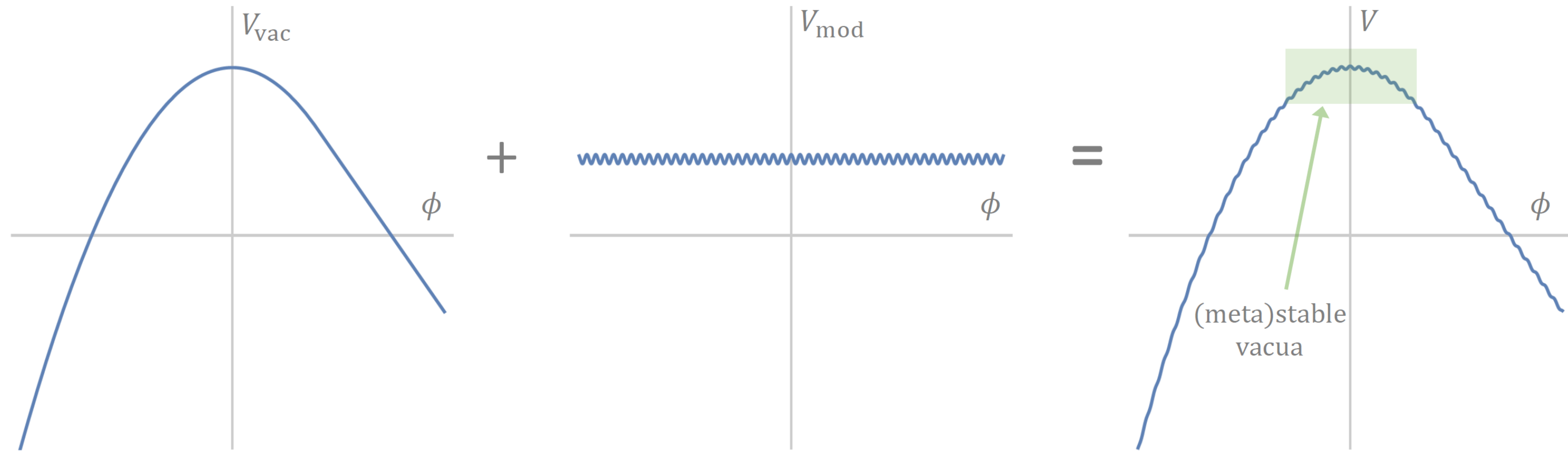}
 \end{center}
 \caption{Illustrative sketch of a mechanism for generating local minima of the $\phi$ field with low weak scale. The first plot on the right depicts the general form of the effective potential/vacuum energy $\Vv(\phi)$ discussed in \Sec{sec:crit}, with a local maximum at the value of $\phi$ corresponding to low weak scale. If we add to this an independent periodic component for the $\phi$ potential $\Vmod(\phi)$ (middle plot), then the resulting total vacuum energy has the form shown in the rightmost plot, with local minima existing sufficiently close to the extremum of the original vacuum energy $\Vv$. }
 \label{fig:wiggles}
 \end{figure}

We have assumed thus far that any potential for $\phi$ is smooth enough to be linearly approximated in the region where $\phi$ scans through the EWSB transition. Since the response of the vacuum energy to symmetry breaking then always yields a concave-down contribution (\Eq{eq:Vpp}), we can never obtain a local minimum for $\Vv$. To avoid this conclusion we introduce new terms to the $\phi$ potential that are sufficiently nonlinear as the weak scale is scanned, so stable minima can exist. In the simplest possible scenario, we add by hand an additional small ``bare'' $\phi$-dependent contribution to the vacuum energy $\Vmod(\phi)$. By construction, this bare potential alone will produce vacua over a range of field space, and has no intrinsic preference for a small EWSB scale. In our geographical analogy, $\Vmod(\phi)$ effectively adds small ridges to the broader mountaintop feature of $\Vv$, as sketched in \Fig{fig:wiggles}. One particularly interesting possibility, discussed in \Sec{sec:lowWeakMinima}, is that these ridges could be large enough to stabilize $\phi$ on the mountaintop where the weak scale is low but not on the slopes with larger Higgs mass.  We will not always require this though when we discuss various mechanisms for predicting a low weak scale in \Sec{sec:mech}.

\subsubsection{Pseudo-Goldstone Boson Realization}

Consider the SM example with vacuum energy given by \Eq{eq:VvSM}. This vacuum energy is generated from both classical and quantum effects, producing a maximum at the field value for which the EWSB scale $v(\phi)$ is the observed small value $\vobs$.  We now want to add a bare potential for $\phi$ which can accommodate minima for a range of $\phi$ values.  A natural candidate to add is a periodic contribution which is invariant under a discrete shift symmetry $\phi \rightarrow \phi + 2\pi f$,
 \eq{
\Vmod(\phi) = M^4  \cos \frac{\phi}{f}
\label{eq:Vper}
} 
where $M$ and $f$ are constants and perturbative control requires $M \lesssim f$. This bare contribution can be highly nonlinear even as \Eq{eq:VvSM} remains an accurate expansion of the non-periodic component $\Vv$ of the vacuum energy. The result for the total vacuum energy including all contributions is illustrated in \Fig{fig:wiggles}. On its own, the periodic potential $\Vmod$ features an infinite discretum of vacua (local minima), all exactly degenerate. Adding $\Vv$ then breaks the degeneracy of the vacua, and can cause them to disappear entirely as its slope becomes large. 

The coexistence of periodic and non-periodic couplings for a field $\phi$, {\it i.e.} axion monodromy, is technically natural and has various ultraviolet completions in both field theory~\cite{Kaloper:2011jz, Dubovsky:2011tu, Yonekura:2014oja, Furuuchi:2015foh} and string theory~\cite{Silverstein:2008sg, McAllister:2008hb, Flauger:2009ab}. Alternately, instead of considering non-periodic couplings for $\phi$, one could imagine that $\phi$ is truly periodic and the linear couplings we invoke are simply expansions for a function of $\phi$ with very large period $F \gg f$. Exponentially large ratios of periods can be dynamically generated through ``clockwork'' models~\cite{Choi:2014rja, Higaki:2014pja, Kaplan:2015fuy}. 

This application of monodromy for a field scanning the Higgs mass may recall the relaxion model~\cite{Graham:2015cka}. However, relaxion models require a periodic $\phi$ potential which turns on in response to EWSB and but is small otherwise, while the periodic potential we invoke is completely independent of EWSB. We further discuss the comparison between our approach and the relaxion model~\cite{Graham:2015cka} in \App{sec:compare}.

The total $\phi$ potential in the SM-like model in the EWSB region is then
\eq{
V(\phi) = \Vv(\phi) + \Vmod(\phi) = -\frac{g^2}{4\lambdaH} \phi^2 + M^4 \cos \phi/f
\label{eq:Vdef}
}
In order for this potential to have local minima at all, the second derivative with respect to $\phi$ must be positive at some points.  Since symmetry breaking gives negative $\Vv'' \sim - g^2 / \lambdaH$, we need 
\eq{
\Vmod'' \sim M^4/f^2 \gtrsim g^2/ \lambdaH,
\label{eq:VppBare}
}
to even have minima in the EWSB region. Barring any fine-tuned cancellation, the resulting mass of $\phi$ around the local minima is $m_\phi^2 \sim \Vmod'' \gtrsim g^2/\lambdaH$. We must also have
\eq{
gf \lesssim \lambdaH \vobs^2,
}
so that the minima scan the Higgs mass finely enough to achieve the observed weak scale.

\subsubsection{Vacua at Observed Weak Scale Only}
\label{sec:lowWeakMinima}

One interesting possibility for this model is that local minima in $\phi$ do not exist when the weak scale is either zero (unbroken EW symmetry phase) or significantly larger than the observed value. In this case, all vacua in the mountaintop region correspond to a parametrically small weak scale. We will make use of this in \Sec{sec:case3} and \Sec{sec:anthroDM}.

Since the slope of $\Vv$ is $\sim g \vobs^2$ in the symmetry preserving region or $\sim g v(\phi)^2$ in the EWSB region, the minima disappear away from low weak scale if
\eq{
\Vmod' \sim M^4/f \lesssim g v_{\rm obs}^2.
\label{eq:smallv}
}
We can use this relation together with \Eq{eq:VppBare} to obtain bounds on the mass and coupling of $\phi$ in this scenario: 
\eq{
m_\phi^2 \sim M^4/f^2 \lesssim M^2 \sim \frac{\Vmod'}{\sqrt{\Vmod''}} \lesssim \sqrt{\lambdaH} \vobs^2.
\label{eq:mphismallv}
}
\eq{
g \gtrsim M^4/f\vobs^2 \sim m_\phi^2 f/\vobs^2 \gtrsim m_\phi^3/\vobs^2
\label{eq:gsmallv}
}

\subsection{Model Constraints}

\subsubsection{Stability Bounds}
\label{sec:stability}

The local minima of this model are of course only metastable, allowing tunneling through the potential barriers to lower minima or to a runaway.

First, let us consider the zero temperature stability of these vacua. In the thin-wall approximation, the rate to tunnel through the nearest potential barrier is
\eq{
\Gamma(\phi) \sim f^4 \exp^{-S(\phi)} \quad \textrm{where} \quad
S(\phi) \sim \frac{55296}{\pi} \times \frac{M^8 f}{g^3 \left[v^2(\phi) - \vobs^2 \right]^3} ,
}
corresponding to a lifetime which can easily be much longer than the age of the Universe for parameters satisfying the conditions outlined previously. 

Second, consider the stability of these vacua at finite temperature. If the Higgs field is thermalized in the early Universe, then the effective potential for $\phi$ will be quite different then, even if $\phi$ itself is not thermalized.   If the Universe reaches temperatures greater than the observed zero temperature EWSB scale, $T \gtrsim v_{\rm obs}$, then a coupling of the form $g \phi |H|^2$ will induce a thermal potential term, $ V_{\rm th}(\phi) \sim g T^2 \phi$. This will destabilize the local minima near the original mountaintop region when the temperature is above some critical value $T_*$ such that 
\eq{
g T_*^2 \sim M^4/f
\label{eq:T*}
}
 
If the zero-temperature vacua are destabilized, $\phi$ rolls away from the mountaintop while the temperature is high. The terminal velocity of the $\phi$ field is $\dot{\phi} \sim V_{\rm th}'(\phi)/H$ where $H \sim T^2/\mPl$ is the Hubble rate. This gives an approximately constant velocity of $\dot{\phi} \sim g \mPl$.  Meanwhile, the age of the Universe at $T \gtrsim T_*$ is $t \sim \mPl/T_*^2$, so that the total field traversal of $\phi$ is $\Delta \phi_T \sim g\mPl^2 /T_*^2$. The region in which the weak scale is of order $\sim v_{\rm obs}$ has size $\Delta \phi_{\rm weak} \sim \lambdaH v_{\rm obs}^2/g$, so in order for $\phi$ not to roll out of this region while the Universe is hot we need $\Delta \phi_T < \Delta \phi_{\rm weak}$, or
\eq{
g^3 \lesssim \lambdaH v_{\rm obs}^2 M^4/(\mPl^2 f) 
\label{eq:gbound}
}
In the particular case where minima only exist for low weak scale, {\it i.e.} \Eq{eq:smallv} holds, this gives $g \lesssim \sqrt{\lambdaH} \vobs^2/\mPl$, or $\Delta \phi_{\rm weak} \gtrsim \sqrt{\lambdaH} \mPl$. In other words, in order for $\phi$ to be able to scan the weak scale by an amount $\gg \vobs^2$, it must scan over a field range $> \mPl$.  

Note however that the constraint \Eq{eq:gbound} is not necessary if the reheating temperature is below the EWSB scale $\vobs$, in which case the thermal correction to the potential is subdominant.

\subsubsection{Observational Constraints}
\label{sec:pheno}

For small excitations around a vacuum, the $\phi$ field appears as a scalar with mass $m_\phi \sim M^2/f$ and a super-renormalizable coupling to the Higgs $g \phi |H|^2$. The effective dimensionless coupling of $\phi$ to the Higgs is $g^2/m_\phi^2$, which using \Eq{eq:VppBare} is bounded $\frac{g^2}{m_\phi^2} \lesssim \lambdaH$; {\it i.e.} $\phi$ is always weakly coupled (barring tuning). For $g^2/m_\phi^2$ sufficiently large there are various probes for $\phi$ depending on the mass, including rare decays and branching ratio modifications at colliders~\cite{OConnell:2006rsp, Batell:2009jf}, and searches for fifth forces and/or equivalence principle violation~\cite{Piazza:2010ye}. However, since the coupling $g$ is allowed to be very weak, even in the scenario of \Sec{sec:lowWeakMinima} where the lower bound \Eq{eq:gsmallv} applies, $\phi$ may be essentially invisible to such probes.

Oscillations of $\phi$ around a local minimum can contribute to the dark matter density. If $\phi$ is initially displaced from a minimum, with typical initial excess energy $\sim M^4$, then the current $\phi$ dark matter density is roughly
\eq{
\rho_\phi \sim \frac{M^4}{m_\phi^{3/2}} \frac{T_{\rm CMB}^3}{M_\text{pl}^{3/2}}
\label{eq:dm}
} 
where $T_{\rm CMB}$ is the current CMB temperature. For appropriate parameters this could constitute the observed total DM density. For very light masses, $m_\phi \lesssim 10^{-6} \eV$, the oscillation in ``fundamental'' parameters such as the electron mass caused by the oscillating background of $\phi$ could potentially be detectable through various experimental techniques~\cite{Arvanitaki:2014faa, Stadnik:2014tta, Arvanitaki:2015iga, Graham:2015ifn, Geraci:2016fva, Arvanitaki:2016fyj, Arvanitaki:2017nhi, Hees:2018fpg, Geraci:2018fax}. This phenomenology is essentially similar to that realized in relaxion models, see e.g.~\cite{Choi:2016luu,Flacke:2016szy,Banerjee:2018xmn,Banerjee:2019epw}, except for the absence of the axion-like coupling $\phi \tilde{G} G$. Accounting for the bound on the effective coupling $g^2/m_\phi^2 < \lambdaH$, current constraints do not rule out any viable parameter space for the model, though some proposed experiments~\cite{Arvanitaki:2014faa, Arvanitaki:2016fyj, Arvanitaki:2017nhi} may have enough sensitivity to begin probing the model.

\subsubsection{Evolution to Distant Vacua}
\label{sec:valley}

Until now we have largely focused on the mountaintop feature of the quantum-corrected vacuum energy $\Vv$ and the conditions to generate stable vacua near it. However, the basic physical constraint that energy is bounded from below implies that at large field values, the vacuum energy must eventually slope back upwards, yielding ``valleys'' in the vacuum energy which are also stable.

Because such valleys are dynamical attractor regions for $\phi$, one should be concerned that in a generic cosmology $\phi$ should be expected to reside there rather than in the mountaintop region with low weak scale. For example, suppose that in different Hubble patches of the early Universe $\phi$ takes on different initial values, with a uniform probability distribution over field space. Then if $\phi$ evolves classically, it will roll to the valley in most regions of the Universe, since the basin of attraction for the valley is much larger than that for the vacua near the mountaintop.

In what follows we will discuss a few different approaches to avoid this outcome. For example, in \Sec{sec:inflation}, regions with $\phi$ settled on the mountaintop will inflate much faster than those in the valley, while in \Sec{sec:anthroDM} we argue that anthropic considerations could disfavor cosmologies which feature rolling to the valley.

\subsection{Late-time Cosmological Constant}
\label{sec:cc}

In our discussion thus far we have computed the potential for an evolving scalar field $\phi$ by recasting it as the vacuum energy as a function of a mass parameter modulated by $\phi$.  However, we have not been concerned with the total, field-independent value of the vacuum energy, {\it i.e.} the cosmological constant (CC).  In fact,  in the model as described thus far, we have treated the total vacuum energy as {\it only} depending on the value of $\phi$, so different $\phi$ vacua necessarily have different cosmological constant. This correlation between the value of the weak scale and the CC  will break down however if there exists some other sector, decoupled from $\phi$ and the Higgs sector, which can scan the CC, {\it i.e.}~through its own landscape of vacua. For most of what follows we will assume that the CC is set to nearly zero independently of the value of $\phi$. An exception will be in \Sec{sec:anthroCC}, where we will consider how the $\phi$-dependent part of the vacuum energy could potentially affect the statistical distribution of the total CC.

\section{Mechanisms for Emergent Criticality}
\label{sec:mech}

By adding a bare potential we have effectively engineered a landscape for the vacuum energy in which the vacua with highest energy are those with a hierarchically small EWSB scale.  In this section we propose several distinct approaches to achieving a cosmology in which our Universe settles into one of these vacua.

\subsection{Inflationary Attractors}
\label{sec:inflation}

Of course, vacua near the mountaintop are not attractor solutions for the evolution of scalar fields, which naturally roll to lower energies. However, the mountaintop can become an attractor in the presence of gravity, since greater vacuum energy induces faster spacetime expansion and thus more spatial volume for observers to form~\cite{Linde:1986fd,Goncharov:1987ir}. In a recent and quite interesting paper \cite{Geller:2018xvz}, Geller, Hochberg and Kuflik proposed precisely such a model in which the EWSB scale is modulated by a dynamical scalar field.  By construction, the potential energy as a function of this field has a maximum coinciding with hierarchically small values of the weak scale.  They then showed that after a long enough period of inflation the volume of the Universe can be dominated by Hubble patches with low weak scale. The model of \Ref{Geller:2018xvz} arranges for a sawtooth-like maximum in the vacuum energy at a low value of the weak scale by having the weak scale control a first-order phase transition in another light field with QCD-axion-like couplings. In this section we will consider the application of essentially the same inflationary attractor mechanism to our class of models in which a smooth local maximum appears in response to EWSB. The comparison of our approach with that of \Ref{Geller:2018xvz} is discussed further in \App{sec:compare}.

In an inflating spacetime, a field $\phi$ may take on different classical values in different Hubble patches, while quantum fluctuations will cause the field values to undergo a ``random walk'' as space expands. At least when considering finite spatial volumes, we may describe the distribution of $\phi$ across different Hubble patches by a function $P(\phi,t)$ satisfying a modified version of the Fokker-Planck (F-P) equation (see {\it e.g.}~\Refs{Linde:1986fd,Goncharov:1987ir,Nakao:1988yi, Nambu:1988je,Starobinsky:1994bd, Vanchurin:1999iv}):
\eq{
\frac{\partial P}{\partial t} = \frac{\partial}{\partial \phi} \left[ \frac{H(\phi)^3}{8\pi^2} \frac{\partial P}{\partial \phi} + \frac{V'(\phi)}{3 H(\phi)} P \right] + 3 H(\phi) P
\label{eq:FP}
}
If we ignore the $3 H(\phi) P$ term, then the above is the ordinary Fokker-Planck equation which describes a normalized probability distribution with diffusion constant $\propto H^3(\phi)$ and a drift term $\propto V'(\phi)/H(\phi)$. The $3 H(\phi) P$ term however causes $P(\phi, t)$ to increase in proportion to the local expansion of space, so that $P(\phi,t)$ becomes an un-normalized, volume-weighted probability distribution for the field value $\phi$ at time $t$.  

We will consider the evolution of $P(\phi, t)$ during a period of slow-roll inflation, starting from some initial distribution $P(\phi, t = 0)$ describing some finite initial volume which will expand until the end of inflation at time $t_{\rm end}$. As usual, the FRW time-slices should be chosen to coincide with surfaces of constant inflaton field value. We will assume that the $\phi$-dependent vacuum energy $V(\phi)$ varies by a small amount compared to the total inflationary energy density. Let us denote the $\phi$-independent component of the inflationary energy density as $\Vinf$, with corresponding Hubble rate $\Hinf = \sqrt{\Vinf/3\mPl^2}$, and take these to be constant throughout inflation for simplicity. The Hubble rate as a function of $\phi$ can be expressed as $H(\phi) = \Hinf + \Delta H(\phi)$ where
\eq{
\Delta H(\phi) \sim \frac{1}{2} \frac{V(\phi)}{\Vinf} \Hinf.
} 
This differential Hubble rate will cause regions of different $V(\phi)$ to expand by different amounts over timescales $\taudiff \sim 1/\Delta H(\phi)$, which occurs after $\Hinf \taudiff \sim \Vinf/V(\phi)$ $e$-foldings of inflation. Although $\taudiff$ may be a relatively long time-scale, this differential expansion can easily become a dominant effect in \Eq{eq:FP} after this time since the other terms are diffusive and probability-conserving. In the following sections we will consider a few different possibilities for the evolution of $P(\phi, t)$ which could lead to regions of low weak scale dominating the late-time volume. 

Regardless of the specifics of how $P(\phi, t)$ evolves, the inflationary background must satisfy a few constraints.  In order to justify integrating out the Higgs field to obtain $\Vv(\phi)$, we must assume that fluctuations of the Higgs are small during inflation, implying
\eq{
\Hinf \lesssim v_{\rm obs}
}
which requires a relatively low inflationary scale $\Vinf \lesssim \Hinf^2 \mPl^2 \sim \left(10^{10} \GeV\right)^4$. As in \Ref{Geller:2018xvz}, we will also assume that this inflationary potential energy is much larger than the vacuum energy scanned by $\phi$, {\it i.e.} $V(\phi) \ll \Vinf$ everywhere, so that $\phi$ has a perturbative effect on the rate and duration of slow-roll inflation, and the Universe is never vacuum energy dominated at the end of inflation.

We will require a long period of inflation in order for a low weak scale to dominate.  However as in \Ref{Geller:2018xvz} we wish to avoid considering eternal inflation and the associated measure problem. This criterion bounds the maximum number of $e$-foldings by  (see {\it e.g.} \Refs{Creminelli:2008es, Dubovsky:2011uy, Graham:2018jyp, Geller:2018xvz})
\eq{
\Ne \lesssim \frac{\mPl^2}{\Hinf^2} \sim \frac{\mPl^4}{\Vinf}
\label{eq:NeNonEternal}
}  
to avoid an eternally inflating regime.

The observable volume of our Universe arose from the expansion of a single Hubble patch or ``progenitor patch'' over the last $\Nobs \sim 50$ e-folds of inflation. In addition to ensuring that $\phi$ most probably takes on a value corresponding to low weak scale in this progenitor patch, we should require that $\phi$ does not fluctuate too much during the last $\Nobs \sim 50$ e-folds of expansion, so that at late times $\phi$ settles into a single local minimum throughout our observed Universe. (If this were not the case, there would be domains of varying weak scale and vacuum energy in the Universe, and any regions of negative vacuum energy could potentially destroy the rest of spacetime~\cite{East:2016anr}.) In the Fokker-Planck language, we can describe the distribution of $\phi$ in the spacetime descendant from a progenitor patch with field value $\phi_0$ by some $P(\phi,t)$ with initial condition $P(\phi,t_i) = \delta(\phi-\phi_0)$. Requiring that it is unlikely to find $\phi$ near a different local minimum after $\Nobs$ e-folds amounts to the condition $P(\phi_0 \pm \pi f, t = t_i + \Nobs \Hinf^{-1}) \lesssim \exp(-3 \Nobs)$ (after normalizing the distribution $P$).

\subsubsection{Case I: $\phi$ Diffuses Freely}
\label{sec:case1}

In some neighborhood of the mountaintop the potential for $\phi$ can be written as 
\eq{
V(\phi) = -\frac{g^2}{4 \lambdaH} \phi^2 + M^4 \cos \phi/f
\label{eq:Vphi}
}
where we have chosen the origin of $\phi$ to correspond to the maximum of the potential at which the EWSB is near its observed value.  To understand the effects of the periodic potential, one can first consider the limit in which differential expansion, described by the $3 H P$ term in \Eq{eq:FP}, can be ignored.  In this case the Fokker-Planck equation has a steady-state solution $P(\phi, t) \propto \exp \left[ -8\pi^2 V(\phi)/3\Hinf^4 \right]$~\cite{Starobinsky:1994bd}. Hence, if $M^4 \gg \Hinf^4$,  then $P(\phi, t)$ will be extremely suppressed away from the local minima of the $\cos \phi/f$ potential, while in the opposite regime $M^4 \ll \Hinf^4$ the cosine barriers can be ignored. We discuss fully the effects of including differential expansion in \App{sec:appendix}, with the same conclusion that the cosine potential is negligible when $M^4 \ll \Hinf^4$.

\begin{figure}
		  \centering
        \subfigure[Case I: $\phi$ diffuses over barriers, $P(\phi,t)$ reaches steady-state with some spread around mountaintop]{
                \includegraphics[width=6cm]{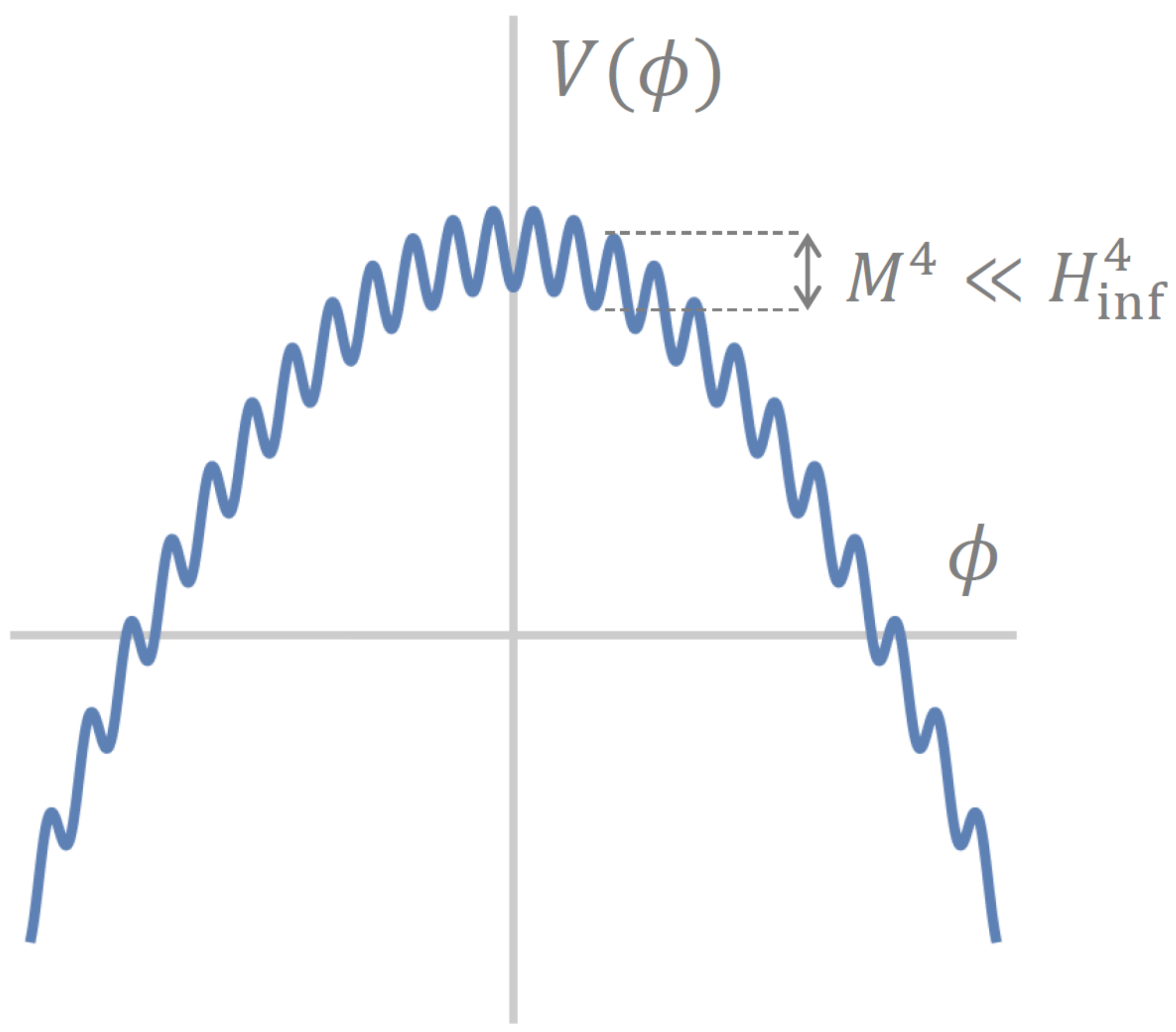}
                \label{fig:case1}
		  }
               \hspace{20mm} 
        \subfigure[ Case II: $P(\phi,t)$ is trapped near local minima, but regions with higher vacuum energy expand faster]{
                \includegraphics[width=6cm]{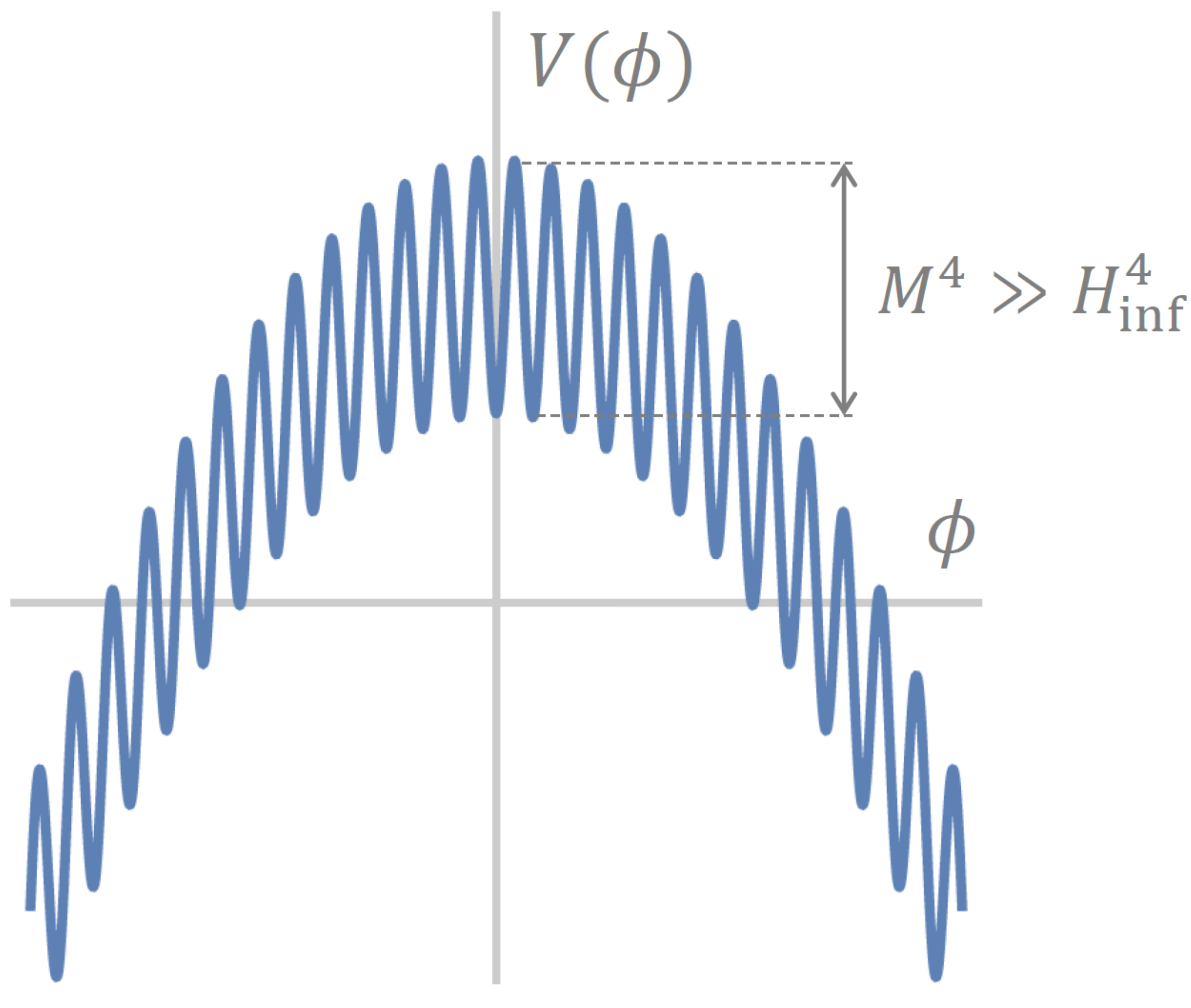}
                \label{fig:case2}
		  }
		
        \subfigure[Case III: $\phi$ is trapped when close to the mountaintop, and otherwise rolls down to a valley with much lower vacuum energy]{
                  \includegraphics[width=15cm]{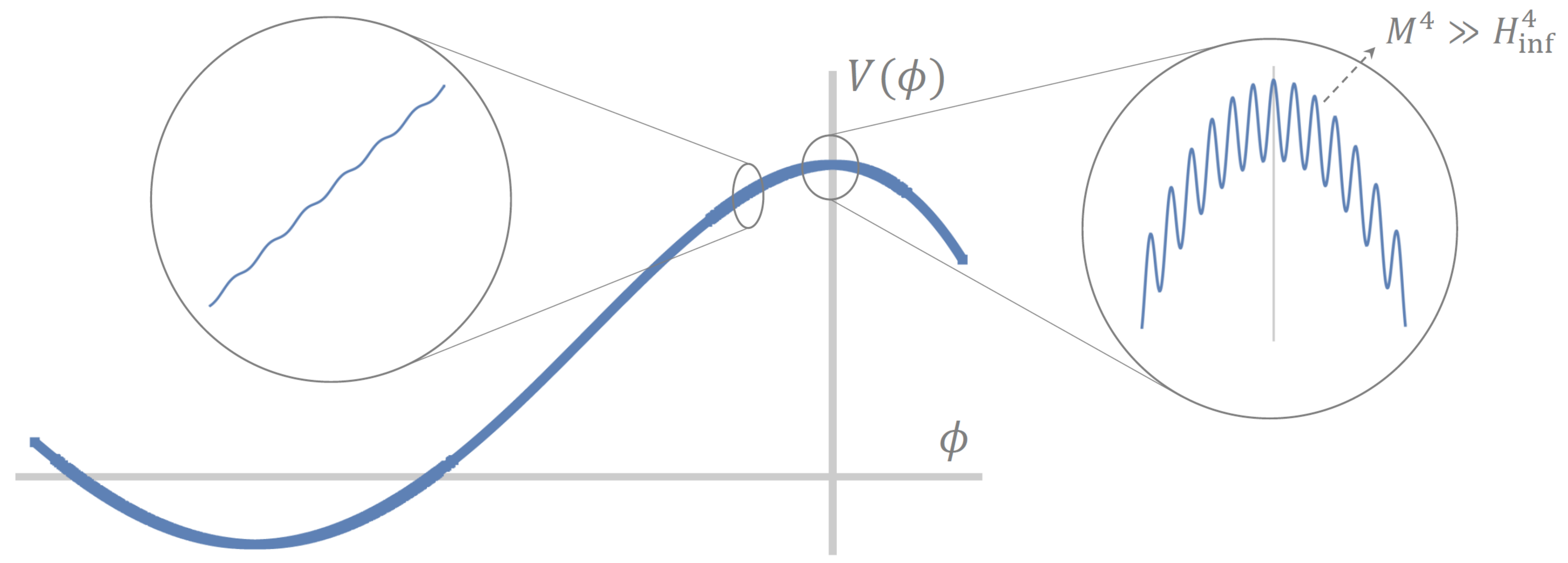}
                \label{fig:case3}
		  }
 \caption{Sketch of the potential $V(\phi)$ for the three different scenarios for the evolution of $P(\phi,t)$.  }
 \label{fig:inflation}
 \end{figure}

Alternatively, when $M^4 \ll \Hinf^4$ the periodic bare potential $M^4 \cos \phi/f$ has negligible effect on the evolution of $P(\phi, t)$, {\it e.g.} see \Fig{fig:case1}. In this case we need only consider the evolution of $P(\phi, t)$ in response to the potential $V(\phi) \sim -g^2 \phi^2/\lambdaH$. In \App{sec:appendix} we discuss the solution of the modified Fokker-Planck equation in \Eq{eq:FP} with a quadratic potential of either sign by finding an eigenvalue decomposition, drawing on the results of Appendix B of \Ref{Graham:2018jyp}. At late times $P(\phi, t)$ will approach a steady-state solution corresponding to the mode with the smallest decay rate, which is a Gaussian centered at $\phi = 0$. 

In the regime of interest to us, the width $\sigma_\phi$ of the Gaussian can be determined by requiring that at large $\phi$ the differential expansion term $3 \Delta H P = -\frac{1}{2} \frac{g^2}{\lambdaH \Hinf}\frac{\phi^2}{\mPl^2} P$ balances against the rolling term $(V'/3H) P \sim - \frac{2}{3}\frac{g^2}{\lambdaH \Hinf} \phi P' \sim \frac{2}{3} \frac{g^2}{\lambdaH \Hinf} \frac{\phi^2}{\sigma_\phi^2} P$ (the other terms in the modified Fokker-Planck equation are subdominant at large $\phi$ if we are in the regime of small $\delta$ as defined in \App{sec:appendix}). This gives $\sigma_\phi = 2 \mPl/\sqrt{3}$.\footnote{This of course requires that the field space of $\phi$ has size $\gtrsim \mPl$. There are various conjectures ({\it e.g.} \Refs{Ooguri:2006in, Blumenhagen:2018nts, Ooguri:2018wrx}) suggesting that scalar field excursions should be sub-Planckian in theories of quantum gravity, including proposals based on scalar analogues of the weak gravity conjecture~\cite{ArkaniHamed:2006dz, Rudelius:2015xta, Brown:2015iha, Heidenreich:2015wga}. However, in \Ref{Saraswat:2016eaz} it was argued that violation of the weak gravity conjecture (for vectors or scalars) can emerge in low-energy EFT without causing any pathological behavior.} In order to predict the correct weak scale, we want $v(\phi)$ to vary by less than $O(1)$ within this width, {\it i.e.} $g \sigma_\phi \lesssim \lambdaH \vobs^2$ or
\eq{
g \lesssim \frac{\lambdaH \vobs^2}{\mPl}
\label{eq:gwidth}
}
As discussed in \App{sec:appendix}, provided that the initial $P(\phi, t = 0)$ is not too localized away from the peak (specifically, that it has at least a Gaussian tail of width $\sim \mPl$ overlapping the region $\phi =0$), it will relax to the steady-state distribution over a typical time $\tau_r \sim \lambdaH \Hinf/g^2$. Requiring inflation to last longer than $\Hinf \tau_r$ gives
\eq{
\Ne \gtrsim \frac{\lambdaH \Hinf^2}{g^2}.
\label{eq:Nediffuse}
}
In order to also satisfy the bound \Eq{eq:NeNonEternal} on the duration of slow-roll inflation, we must have 
\eq{
g^2/\lambdaH \gtrsim \Hinf^4/\mPl^2
\label{eq:deltaBound}
}

Therefore when the above conditions on $g, \Ne$ and $P(\phi,t=0)$  are satisfied, the distribution $P(\phi,t)$ at the end of inflation will be dominated by the mountaintop where the weak scale is low. 

As discussed above we should also verify that $\phi$ typically occupies a single local minimum in the volume that will form our observable Universe, which forms from a single progenitor patch after the last $\Nobs \sim 50$ e-folds of expansion. Let us require that this time is much less than the dynamical relaxation time, i.e. $\Nobs \ll \lambda \Hinf^2/g^2$. Then given a progenitor patch with some field value $\phi_0$, the field values in its descendant Hubble patches will be determined by a simple random walk (pure diffusion in terms of the F-P equation), giving a distribution of $\phi$ that is Gaussian with variance $\sim \Nobs \Hinf^2 $. The requirement that there is low probability to find $\phi$ displaced by order $f$ in the resulting volume translates to $\exp[-f^2/(\Nobs \Hinf^2 )] \lesssim \exp[-3 \Nobs]$, or $f \gtrsim \Nobs \Hinf $, which is easily satisfied.   
 
\subsubsection{Case II: $\phi$ Locally Trapped}
\label{sec:case2}

The opposite regime to consider for $V(\phi)$ in \Eq{eq:Vphi} is the case where the periodic potential is large enough to strongly suppress diffusion of $\phi$, which occurs when $M^4 \gtrsim \Hinf^4$ (\Fig{fig:case2}). As discussed above, without differential expansion this would simply cause $P(\phi,t)$ to accumulate around local minima of the potential with an exponentially suppressed rate to diffuse through the barriers to other regions. Including the effect of differential expansion, $P(\phi,t)$ still localizes around minima provided that $f < \mPl$ (see \App{sec:appendix}), though in the opposite regime $P(\phi,t)$ will instead localize around maxima of the potential, as noted in \Ref{Graham:2018jyp}. We will focus on $f < \mPl$ so as to avoid $\phi$ potentially exiting the region of low weak scale when it rolls classically after the end of inflation.

Since the probability density $P(\phi,t)$ can neither roll over or diffuse through the periodic barriers, it is clear that the local minima with maximal vacuum energy will come to dominate due to differential expansion, provided they are populated with some non-zero initial density. The difference in Hubble rate between a point with the correct weak scale and regions with $O(1)$ higher weak scale is $\Delta H \sim \Hinf \frac{\lambdaH \vobs^4}{\Vinf}$, so a low weak scale becomes exponentially favored on the timescale $\taudiff \sim 1/\Delta H$ which can be realized for 
\eq{
\Ne \gtrsim \Hinf \taudiff \sim \frac{\Vinf}{\lambdaH \vobs^4}.
\label{eq:NeDiffExp}
} 
Note that the above condition is also necessary in the previous case, where it follows from \Eq{eq:Nediffuse} and \Eq{eq:gwidth}. In order for regions near $\phi =0$ to actually dominate after the time given above, their initial population should not be too suppressed. Using similar reasoning as discussed in \App{sec:appendix}, we find that it is sufficient for the initial density to be of the form $P(\phi, t) \sim \exp [-(\phi-\phi_i)^2/\sigma_0^2]$ with $\sigma_0 > \lambda \vobs^2/g$ and $\phi_i$ arbitrary.

The above discussion assumes that the rate of transitions between vacua, which occur through the Hawking-Moss instanton~\cite{Hawking:1981fz} with suppression $\sim \exp(-16\pi^2 M^4/9 \Hinf^4)$, is slower than the timescale \Eq{eq:NeDiffExp} over which weak-scale vacua dominate. This condition also implies that if $\phi$ began in a particular vacuum in some progenitor Hubble patch, then there is an extremely suppressed probability to explore any other vacua in the $\Nobs \approx 50$ e-folds of expansion that produce our observed Universe.

\subsubsection{Case III: $\phi$ Populates only Mountaintop and Valley}
\label{sec:case3}

The lower bound on $\Ne$ in \Eq{eq:NeDiffExp} is quite large compared to the $\sim 60$ $e$-folds required by the horizon problem, since $\Vinf$ should be taken to be much larger than the weak scale if $V(\phi)$ is to be everywhere treated as a small perturbation of the inflationary energy density. While this is somewhat peculiar, it is not necessarily inconsistent, as this lower bound can be well below the maximal $\Ne$ that can be realized for non-eternal slow-roll inflation, \Eq{eq:NeNonEternal}. However, it is interesting to ask whether it is possible for a low weak scale to dominate within a shorter period of inflation. This could be possible if $\phi$ can only populate two regions during inflation, the mountaintop where the weak scale is low and a valley which forms when the potential starts to turn up again due to some other dynamics of $\phi$ (illustrated in \Fig{fig:case3}). Let us denote the value of the weak scale where the expansion $V(\phi) = -\lambdaH v(\phi)^4/4$ starts to break down as $v_{\rm max}$. Then the difference in Hubble rate between the mountaintop and the valley is at least $\left( \lambdaH v_{\rm max}^4/\Vinf\right) \Hinf$ and the necessary number of $e$-folds becomes
\eq{
\Ne \gtrsim  \frac{\Vinf}{\lambdaH \vmax^4}.
\label{eq:NeDiffExp2}
}
which could conceivably be $O(100)$. 

Let us outline the conditions necessary to achieve such a scenario. As in \Sec{sec:lowWeakMinima}, we will require that the periodic potential is small enough that it only creates minima in some neighborhood of $\phi = 0$ where the weak scale is low; however it must also be tall enough that $\phi$ does not diffuse through it, giving 
\eq{
 \Hinf^4  < M^4 < g \vobs^2 f.
\label{eq:groll}
}
With these conditions, any initial probability density $P(\phi, t)$ very close to $\phi = 0$ will be trapped by the barriers with an exponentially suppressed rate to leak out, while in regions where $v(\phi) \gtrsim \vobs$, $\phi$ feels a classical potential  $\sim -g^2 \phi^2/\lambdaH$ with no barriers. Let us for now simply consider the classical evolution of $\phi$ in this region. As $\phi$ rolls down the quadratic it grows exponentially on the timescale $\tau \sim \lambdaH \Hinf/g^2$, so the number of $e$-folds required for it to settle to a distant valley is
\eq{
\Ne \gtrsim \lambdaH \Hinf^2/g^2.
\label{eq:NeRoll}
}
This constraint is trivial if $g > \sqrt{\lambdaH} \Hinf$, in which case $\phi$ rapidly rolls to the minimum at the valley and oscillates, losing most of its kinetic energy after just a few $e$-folds. 

If we account for diffusion of $\phi$, but neglect differential expansion (as is appropriate if $\phi$ rolls over a range less than $\mPl$), then the steady-state distribution is $P(\phi, t) \propto \exp \left[ -8\pi^2 V(\phi)/3\Hinf^4 \right]$. Since $v_{\rm max}^4 \gg \vobs^4 \gtrsim \Hinf^4$, the distribution away from the minima on the mountaintop will concentrate in regions with $v(\phi) \sim v_{\rm max}$. If differential expansion is important, {\it i.e.} the range for $\phi$ is greater than $\mPl$, then on the slope of the mountaintop the analysis of \App{sec:appendix} is applicable , where we find that around an inverted quadratic potential $P(\phi, t)$ decays at a rate $\sim -g^2/\lambdaH \Hinf$ or greater, which is the same timescale found above. 

Therefore with the above conditions, an initial approximately uniform distribution of $P(\phi,t)$ will evolve to include one population where $\phi$ is trapped in a stable region near the mountaintop ({\it i.e.} low weak scale) and another population with $\phi$ rolling away to a distant minima of $V(\phi)$ and settling there. Even if the initial field space in which $\phi$ is stable and gives the correct weak scale is very small, it will exponentially expand and dominate on timescales greater than that of \Eqs{eq:NeDiffExp2}{eq:NeRoll}.

 In order for this scenario to actually produce a low weak scale with only the modest duration of inflation given by \Eq{eq:NeDiffExp2}, the reheating temperature of the Standard Model must be below the weak scale. If this were not the case, then as discussed in \Sec{sec:stability} there is an upper bound on $g$ given by \Eq{eq:gbound} in order for $\phi$ not to roll away from the zero-temperature minimum.  When combined with \Eq{eq:groll} this gives $g^2 < \lambdaH \vobs^4/\mPl^2$. Then \Eq{eq:NeRoll} implies that inflation must last for at least $\Ne \gtrsim \Hinf^2 \mPl^2/\lambdaH \vobs^4 \sim \Vinf/\lambdaH \vobs^4$ e-folds, i.e. no improvement over the bound of \Eq{eq:NeDiffExp} which applied for the preivous two cases.

\subsection{Anthropic Arguments}
\label{sec:anthropic}

In the previous section we described how the Universe could dynamically evolve to be dominated by Hubble patches with maximal vacuum energy after a long period of low-scale inflation. In the absence of these dynamics, {\it e.g.} if inflation does not last sufficiently long, one would instead expect $\phi$ to populate its field space relatively uniformly in different Hubble patches.  In this case we must ask why we happen to live in a Universe with maximal vacuum energy with respect to $\phi$. In this section we propose anthropic arguments for this condition, based on the late-time dark matter density (\Sec{sec:anthroDM}) and cosmological constant (\Sec{sec:anthroCC}).

It should be emphasized that we will be discussing anthropic motivations for living near a maximum of the vacuum energy, \emph{not} for a low weak scale directly. This approach is therefore distinct from the ``atomic principle'' argument for anthropic selection of the weak scale~\cite{Agrawal:1997gf}. In particular, our arguments will not depend on the precise values of SM parameters such as the Yukawa couplings.   
 
\subsubsection{Avoiding the Valley}
\label{sec:anthroDM}

Recall from \Sec{sec:lowWeakMinima} that the periodic bare potential may create stable vacua on the mountaintop only in the region where the weak scale is low, provided that \Eq{eq:smallv} holds, i.e.
\eq{
\Vmod' \sim M^4/f \lesssim g v_{\rm obs}^2.
} 
With this condition, any point near the mountaintop with the wrong value of the weak scale will be unstable. However, as noted in \Sec{sec:valley}, the vacuum energy $\Vv$ is ultimately expected to have some minimum or ``valley'' around some point $\phi_{\rm min}$ far away from the mountaintop feature at $\phi_{\rm max}$ which we have focused on. This valley is the actual classical attractor of $\phi$, absent the effects of a long period of low-scale inflation as discussed in the previous section.

\begin{figure}
\begin{center}
  \includegraphics[width=18cm]{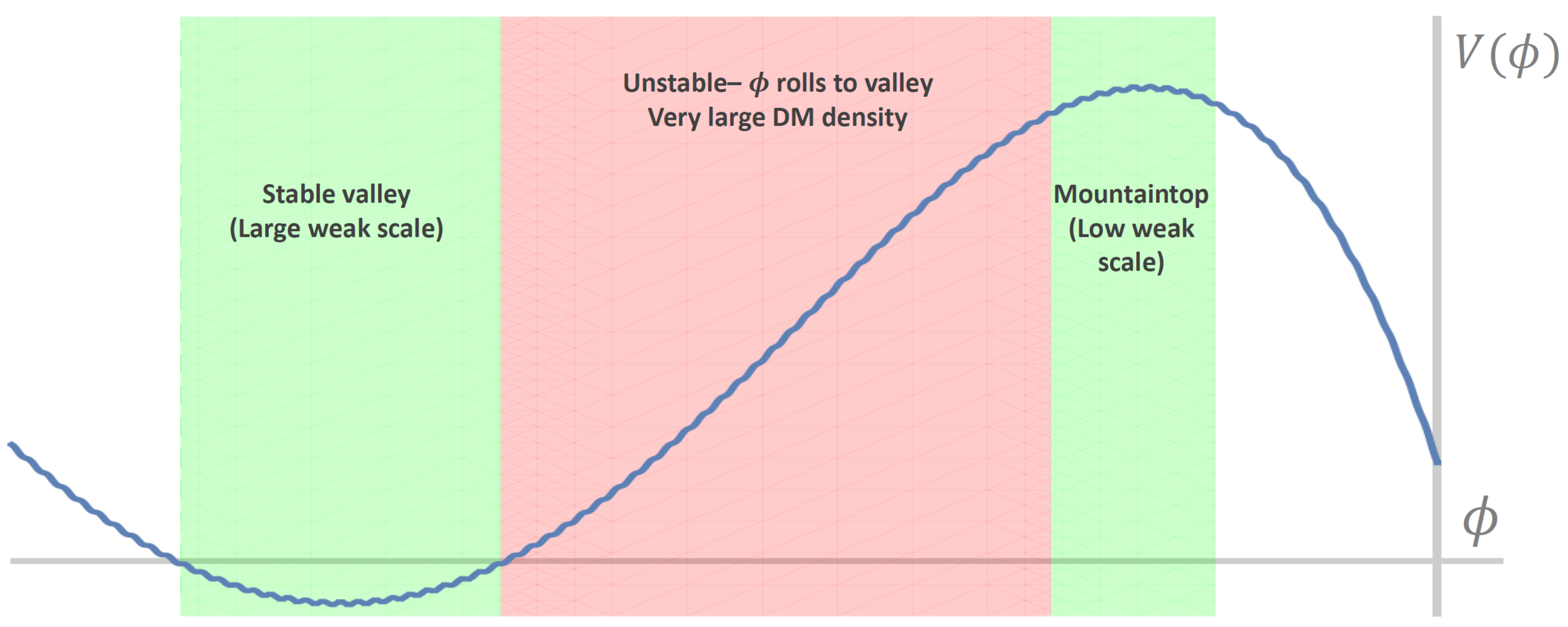}
 \end{center}
 \caption{Sketch of a potential for $\phi$ as considered in \Sec{sec:anthroDM}, featuring both ``mountaintop'' and ``valley'' regions.  With the inclusion of a periodic bare potental, stable vacua may exist near the mountaintop and the valley (green shaded regions) but not on the ``slope'' in between (red shaded region). While it is generic for $\phi$ to begin somewhere in the large slope region, its resulting evolution can produce extremely large dark matter densities, so that only cosmologies with $\phi$ starting on the mountaintop or in the valley may be anthropically viable.
}
 \label{fig:anthroDM}
 \end{figure}

The mass of $\phi$ around the valley (ignoring the periodic potential) is expected to be similar to the curvature at the mountaintop $\sim g^2$ by naturalness. Then there will be a similar number of stable local minima within the valley as around the mountaintop. But given an initial uniform distribution for $\phi$ over different Hubble patches, most regions will start with $\phi$ on the unstable ``slope,'' such that $\phi$ rolls towards the valley and oscillates about it, eventually settling to one of the minima in the valley where the Higgs mass is large. This situation is illustrated in \Fig{fig:anthroDM}.

However, in Hubble patches with this typical evolution, the oscillation of $\phi$ gives a very large dark matter density compared to Hubble patches settled on the mountaintop, which could be anthropically catastrophic. The DM density in the case that $\phi$ begins trapped near a local minimum on the mountaintop was discussed in \Sec{sec:pheno}.  In this case $\phi$ starts with typical excess energy $\sim M^4$ relative to the minimum and oscillates with frequency $m_\phi \sim M^2/f \gtrsim g/\sqrt{\lambdaH}$. In contrast, if $\phi$ rolls down the slope and oscillates across the valley, it will start with a very large energy relative to the minima in the valley, at least $\gg v_{\rm obs}^4 \gg M^4$, and will oscillate at a lower frequency $\sim g \ll m_\phi$. The combination of large initial energy density and lower oscillation frequency ({\it i.e.} more time before the energy starts decaying as $a^{-3}$) implies that the late-time dark matter abundance (more precisely the dark-matter-to-photon ratio) is always parametrically larger in the case of $\phi$ rolling down the slope compared to staying on the mountaintop.

Various considerations for anthropic bounds on the dark matter abundance ({\it e.g.}~\Refs{Hellerman:2005yi, Tegmark:2005dy}) suggest upper bounds of $10^2 - 10^5$ times the observed value. So in models in which the dark matter density varies continuously with some parameter, such as the QCD axion with large $f_a$, anthropic considerations usually do not suffice to explain the observed abundance. However, in our model the dark matter abundance is sharply discontinuous depending on whether or not $\phi$ is trapped near a local minimum, so anthropic arguments can have significant exclusionary force even if the upper bound on the dark matter abundance is taken to be quite large.  

In summary, for appropriate choice of parameters it is possible that the only anthropically viable cosmologies are those with $\phi$ oscillating around a single local minimum on the mountaintop or valley. Since these we expect a similar number of vacua in the two regions, it is at least typical for viable Hubble patches to have a low weak scale (live on the mountaintop).

\subsubsection{Seeking the Mountaintop}
\label{sec:anthroCC}

As we have emphasized, any model which scans the Higgs mass necessarily scans the vacuum energy, {\it i.e.}~the cosmological constant (CC). Thus the smallness of the Higgs mass and the CC are potentially intertwined problems.   
The CC problem itself is often addressed by invoking anthropic selection from a large landscape of vacua. An intriguing possibility is that within some particular landscapes, small values of the CC could be correlated with a low weak scale, so that the same anthropic selection addresses both problems. A model of this type, in which the number density of vacua is enhanced when the weak scale is low, was proposed in \Ref{Arvanitaki:2016xds}. In this section we discuss another approach to achieving a correlation between a small CC and a small Higgs mass, using the models we have discussed where the vacuum energy is maximized for low weak scale.          

The anthropic argument for the smallness of the CC~\cite{Weinberg:1987dv} relies on the observation that cosmological structure formation could not have occurred if the magnitude of the CC were larger than some bound $\rho_{\rm anth}$ (assuming all other parameters, {\it e.g.}~the primordial density fluctuations, are kept fixed). In order to actually realize such a small CC, there must be some landscape of vacua with a statistical distribution of CC values described by some number density $n(\rho)$, such that there exists at least one vacuum with sufficiently small CC, {\it i.e.} $n(\rho = 0) \rho_{\rm anth} \gtrsim 1$. To preserve the anthropic argument in its simplest form, other low-energy parameters such as the expected dark matter density should not vary significantly within this landscape.

\begin{figure}
\begin{center}
  \includegraphics[width=17cm]{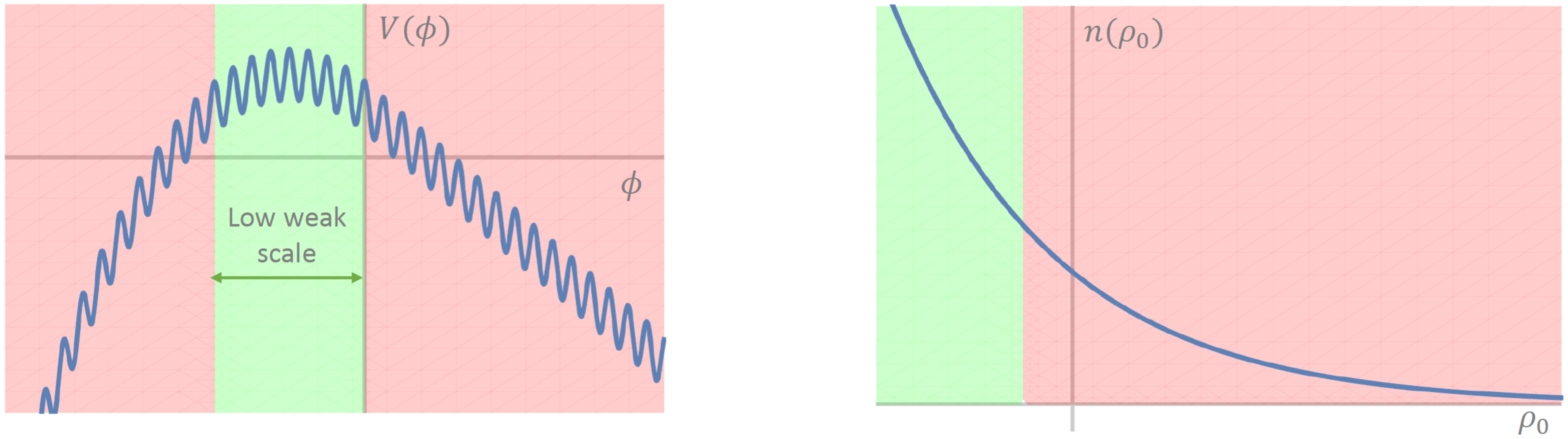}
 \end{center}
 \caption{Left: The blue curve depicts the form of the effective potential for $\phi$ as discussed in \Sec{sec:minima}. The green shaded region indicates the set of vacua yielding an acceptable low weak scale, while the red regions represent vacua with unbroken EWSB or too large a weak scale. Right: A sketch of a possible form for the distribution $n(\rho_0)$ discussed in the text. The green region here indicates the range of $\rho_0$ which can cancel the potential at the minima in the green region of the left figure (low weak scale) to give a total CC $\sim 0$ as required anthropically. Similarly the $\Lambda_0$ values in the red region can cancel the total CC when the weak scale is large. If the distribution $n(\rho_0)$ falls off rapidly enough, then the CC is most likely to be canceled to zero for vacua in the green region, with low weak scale.
 }
 \label{fig:CC}
 \end{figure} 
 
Consider taking a ``tensor product'' of such a landscape of CC values with the landscape of Higgs masses as discussed in \Sec{sec:minima}. (Unlike the previous section let us not assume that stable vacua only exist near the mountaintop where the weak scale is low.) The total cosmological constant will then be the sum of a contribution from $\Vv(\phi)$ which is correlated with the Higgs mass, plus a contribution $\rho_0$ from the rest of the landscape, which is distributed according to some $n(\rho_0)$ as above (independent of the Higgs mass). An anthropically favorable vacuum (small total CC) will then have $\rho_0 \approx - \Vv(\phi)$, so for fixed $\phi$ the number density of viable vacua is $n\left[-\Vv(\phi)\right]$. This scenario is illustrated in \Fig{fig:CC}. Anthropic selection of the CC would simultaneously solve the hierarchy problem if there are a greater total number of viable vacua for small Higgs mass (green regions in the plots of \Fig{fig:CC}) than for large Higgs mass (red regions). The difference in vacuum energy between the maximum of the potential and the start of the red regions is of order $\lambdaH v_{\rm obs}^4$. Since there are many more $\phi$ minima in the red regions than the green, selecting a small Higgs mass requires that $n(\rho_0)$ fall off rapidly as $\rho_0$ increases by $\sim \lambdaH \vobs^4$.

It remains to be determined whether such statistical behavior can be realized in a landscape arising from a ultraviolet complete theory such as string theory. If we consider the simplest ansatze for $n(\rho_0)$, an exponential or Gaussian distribution centered around some value $\Lambda_\textrm{UV}^4$ of order the cutoff, then requiring that there exist vacua with $\Lambda \sim 0$ implies that the total number of vacua is super-exponentially large, {\it e.g.}~scaling as $\exp\left(\frac{\Lambda_\textrm{UV}^4}{\lambdaH \vobs^4}\right)$. This may be difficult or even impossible to realize in a consistent ultraviolet completion. One could however imagine other (more complicated) forms for $n(\rho_0)$. For example, if there are two sectors, one which scans the CC with a Gaussian distribution with standard deviation $\sim \lambdaH \vobs^4$, and another which scans it uniformly in steps of order $\lambdaH \vobs^4$, so that the neighborhood of $\Lambda \approx 0$ could be populated with the appropriate distribution locally. 

As is the case with the standard anthropic approach to the CC problem, the existence and nature of the necessary landscape is not accessible to low-energy experiment. However, the solution to the hierarchy problem in this scenario still relies on $\Vv(\phi)$ naturally having an extremum for low weak scale, which is testable and falsifiable, {\it e.g.}~as will be discussed for the MSSM example of \Sec{sec:SUSY}.

\subsection{Exotic Mechanisms}

In this section we discuss exotic approaches by which the Universe could settle to the maximum of the quantum-corrected vacuum energy $\Vv$.

One way to realize the desired behavior for the classical zero-mode of the $\phi$ field would be to reverse the sign of its kinetic term in the Lagrangian, {\it i.e.}~make $\phi$ a ghost. The local maximum of vacuum energy is then a stable attractor solution for the zero-mode of $\phi$. Of course, ghost fields are known to have catastrophic pathologies---classically, the vacuum is unstable to decay into negative energy states, and quantum-mechanically there is violation of unitarity due to negative norm states. It is unknown whether or how these pathologies can be resolved, and we will not address those valid concerns here.  Nevertheless, we find it amusing that the hierarchy problem in the SM can be solved by the addition of a single ghost field. 

Another exotic approach is the following remarkably minimal addition to the SM which does not require the addition of {\it any} new degrees of freedom.  To begin, let us define 
\eq{
\Omega =\frac{1}{2} \int d^4 x \,  H^2
}
which is the integral over spacetime of the scalar mass operator.  Now consider the following exotic non-local contribution to the action,
\eq{
S_{\rm exotic} = \frac{1}{2}M^4 \Omega^2. \label{eq:S_exotic}
}
We can rewrite this contribution by integrating in an auxiliary parameter which with foresight we dub $\mu^2$, so the this contribution is equivalent to 
\eq{
S_{\rm exotic} = -\mu^2 \, \Omega - \frac{\mu^4}{2M^4} .
}
The first term is precisely a contribution to the Higgs mass. Let us for simplicity take the bare Higgs mass parameter $\mu_0^2$ to be zero at tree-level. When we compute quantum corrections in this theory, we obtain by analogy with \Eq{eq:Vq} 
\eq{
V+ \Delta V+ V_{\rm exotic} =  - \Delta M^2 \mu^2  + \frac{1}{2}\left(\mu^2 + \Delta \mu^2  \right)  H^2 + \frac{\mu^4}{2L^4M^4}\ldots, 
}
where $\Delta M^2$ and $\Delta m^2$ are defined as in \Eq{eq:musq} and $L^4$ is the volume of spacetime which appears because the auxiliary field is a spacetime zero mode.  Again, we can apply the mass shift in \Eq{eq:mass_shift}, but now we see that this induces new $\mu^2$-dependent terms that scale as $\mu^2 \Delta \mu^2 / L^4 M^4$.  In order for these not to ruin the structure of the potential, we require these terms to be sufficiently small, so
\eq{
M^4 \gtrsim \frac{\Delta \mu^2}{\mu^2 L^4},
}
which gives a bound on the size of the original exotic operator.

If this condition is satisfied, then the auxiliary field $\mu^2$ has a potential of the same form as that considered for the dynamical field $\phi$, $\Vv$.  Crucially, since $\mu^2$ is non-dynamical, its equations of motion can set it to the maximum of $\Vv$ simply because it is an extremum.  While this behavior may seem ghost-like, bear in mind that this auxiliary field is fictitious, and this is a roundabout way of computing the physical effect of the exotic operator in \Eq{eq:S_exotic}. Note that a similar mechanism has been discussed previously in the context of relaxing the CC using wormholes in Euclidean quantum gravity \cite{Coleman:1988tj}.

\section{Explicit SUSY Model and Experimental Consequences}
\label{sec:SUSY}

Here we consider a realistic implementation of our proposal within an extension of the MSSM featuring a scanning field $\phi$. SUSY is a favorable setting for this approach to the hierarchy problem for the usual reasons: it explicitly cuts off quadratically divergent loops and renders the weak scale calculable in terms of the SUSY breaking masses, which are generated at some high SUSY messenger scale $\Lambda$. However, as discussed in \Sec{sec:hierarchy}, adding the principle of scanning to an extremum of the vacuum energy will allow the superpartners to be considerably heavier than in the MSSM, without invoking any fine-tuning. 

\subsection{Implementation in Extension of MSSM}
\label{sec:mssm}
Let us introduce $\phi$ as a chiral superfield which couples to the MSSM Higgs doublets in the superpotential,
 \eq{
 W = \mu(\phi) H_u H_d +\cdots
 } 
which defines an effective $\phi$-dependent Higgsino mass parameter,  $\mu(\phi) = \mu_0 + y_\phi \phi$. In the limit of unbroken SUSY, $\phi$ will have zero bare potential even after quantum corrections. For brevity, we will refer to this model here as the $\phi$MSSM.  This model has the same field content as the NMSSM, {\it i.e.}~a single additional neutral chiral superfield. However we are interested in a very different parameter regime than is usually considered for the NMSSM.  For example we do not introduce any other superpotential terms for $\phi$ such as a mass term (which is a technically natural choice due to SUSY non-renormalization). 

Regarding SUSY breaking, we assume as usual that it is communicated to the MSSM fields via some messenger sector at a high scale $\Lambda$.  The soft parameters are then evolved via the renormalization group (RG) down from that scale to determine their low energy values.   Moreover, we will assume that $\phi$ does not couple directly to the messenger sector, so its SUSY-breaking potential and couplings are zero at the high scale $\Lambda$. Nevertheless, starting from this boundary condition, RG flow will generate a SUSY-breaking potential for $\phi$, $\Vvq(\phi)$, as well as a soft trilinear scalar couplings to the Higgs: 
\eq{
V_{\rm soft} = \Vvq(\phi) + b(\phi) H_u H_d + \textrm{c.c.} + \cdots
}
where $b(\phi) \equiv b_0 + a_\phi \phi$ and $a_\phi$ is the trilinear scalar coupling.

To simplify the discussion, we will assume that the coupling $y_\phi$ is weak, so loops of $\phi$ or its fermionic superpartner are negligible. Then for the purposes of RG flow, $\phi$ can be treated as a background field, so the RG evolution of these effective parameters are exactly the same as in the MSSM.  That is, the beta functions for $\mu(\phi) = \mu_0 + y_\phi \phi$ and $b(\phi) = b_0 + a_\phi \phi$ obey the same RG equations as $\mu$ and $b$ in the MSSM. From this we can obtain the beta functions for the couplings $y_\phi$ and $a_\phi$,
\begin{align}
16\pi^2 \frac{dy_\phi}{d\log Q}  &= y_\phi \left(3 y_t^2 + 3 y_b^2 + y_\tau^2 -3 g_2^2 - \frac{3}{5} g_1^2 \right) \\
 16\pi^2 \frac{da_\phi}{d\log Q}  &= a_\phi \left(3 y_t^2 + 3 y_b^2 + y_\tau^2 -3 g_2^2 - \frac{3}{5} g_1^2 \right) \nonumber \\
&\qquad + y_\phi \left(6 a_t y_t + 6 a_b y_b + 2 a_\tau y_\tau + 6 g_2^2 M_2 + \frac{6}{5} g_1^2 M_1 \right)
\end{align}
where $Q$ is the renormalization scale, and we employ the notation of \Ref{Martin:1997ns} for the MSSM parameters. Note that no other couplings of $\phi$ are generated in this limit.  This is a reflection of the fact that $\mu$ and $b$ do not enter into the RG evolution of any other MSSM parameters except the CC because they are super-renormalizable and chiral symmetry breaking. 

To begin, we compute the classical vacuum energy in the $\phi$MSSM arising from EWSB, $\Vvc(\phi)$.
The scalar potential for the Higgs bosons and $\phi$ is 
\eq{
V(H_u,H_d,\phi) &=  (|\mu(\phi)|^2 + m_{H_u}^2) |H_u|^2 + (|\mu(\phi)|^2 + m_{H_d}^2) |H_d|^2 \\ 
& \qquad - [ b(\phi) H_u \cdot H_d + \text{c.c.} ]   \\
& \qquad + \frac{1}{8}\left(g_1^2 + g_2^2\right) \left(|H_u|^2 -|H_d|^2 \right)^2  + \frac{1}{2} g_2^2 |H_u^\dagger \cdot H_d |^2
}
As usual, when computing the Higgs VEVs, the parameters here should be evaluated at a low scale of order the superpartner masses. From the arguments of \Sec{sec:classical}, we know that the vacuum energy generated by integrating out the Higgs doublets will be zero for some range of $\phi$ and then turn on at some critical value where EWSB occurs.  The vacuum energy will be concave-down at this boundary. Therefore, provided that the bare potential $\Vvq(\phi)$ has an appropriate slope near this critical point, there will be a maximum in the total vacuum energy corresponding to a small value for the Higgs VEVs, as depicted in \Fig{fig:plateau}.  

When $|\mu(\phi)|^2 < \frac{1}{2} \left[  \sqrt{4 b(\phi)^2 + \left( m_{H_d}^2-m_{H_u}^2 \right)^2} -m_{H_d}^2 - m_{H_u}^2 \right]$, the above potential develops a tachyonic instability and the Higgs fields acquire VEVs. As usual in the MSSM, $\tan \beta \equiv v_u/v_d$ and the $Z$ boson mass satisfy\footnote{For simplicity, we assume throughout that $m_{H_u}^2 < m_{H_d}^2$, so $\cos 2 \beta$ is negative.}
 \eq{
\sin 2 \beta(\phi)  &= \frac{2 b(\phi)}{m_{H_u}^2+m_{H_d}^2 - 2 |\mu(\phi)|^2} \\ 
m_Z^2(\phi) &= -\frac{m_{H_d}^2-m_{H_u}^2}{\cos2 \beta(\phi) }-m_{H_u}^2-m_{H_d}^2-2 |\mu(\phi)|^2
} 
Plugging these expressions into the tree-level Higgs potential yields the effective classical vacuum energy in the EWSB vacuum,
 \eq{
\Vvc(\phi) = \frac{1}{2} \frac{1}{g_1^2+g_2^2} m_Z^4(\phi) \cos 2 \beta(\phi) <0,
 \label{eq:VeffMSSM}
 }
which as expected from general arguments is negative. Note that at the critical boundary where the $Z$ boson mass vanishes, both the value and first derivative of $\Vvc$ are zero, also as expected.

Next, we compute the $\phi$-dependent quantum corrections to the vacuum energy, $\Vvq(\phi)$.  As noted earlier, we assume that $\phi$ has no bare potential at the SUSY messenger scale $\Lambda$.  At low energies the bare potential is then generated purely from RG running within the $\phi$MSSM. Just as with the running of $y_\phi$ and $b_\phi$, we can easily extract the RGE for the potential for $\phi$ in the small $y_\phi$ limit by considering the RGE for the CC in the MSSM (see {\it e.g.}~\Ref{Martin:2001vx}), treating $\phi$ as a background field. This gives

\begin{align}
\frac{d}{d\log Q} \Vvq(\phi) = \frac{1}{4\pi^2} \left[\left(m_{H_u}^2+m_{H_d}^2\right) |\mu(\phi)|^2 + b\left(\phi\right)^2 \right]+ \cdots
\label{eq:betaV}
\end{align}  
where the ellipses denote terms independent of $\phi$.  The RG-improved result for the potential is then obtained by running the above equation, along with the RGEs for all MSSM parameters, starting from the mediator scale $\Lambda$.  The running masses of the stops, gauginos, etc. enter into the result for $\Vvq$ through the RGEs of the Higgs soft parameters.  As usual, in the absence of fine-tuning the Higgs soft masses are expected to be not far below the stop and gaugino masses, particularly if the messenger scale $\Lambda$ is high so that there is considerable RG running.     

The condition that the total vacuum energy is extremized with respect to $\phi$ can be expressed as 
 \eq{
\Vvc'(\phi) + \Vvq'(\phi) = 0 \label{eq:dVdmu}
 }

Given the SUSY breaking parameters at the mediation scale $\Lambda$, one can RG evolve to obtain the quantum corrected vacuum energy and the Higgs potential parameters at a low scale, and then compute the field value $\phi_{\rm max}$ where the vacuum energy is maximized.   This condition then fixes the effective MSSM parameters and in turn the scale of EWSB.  

\begin{figure}
\begin{center}
  \includegraphics[width=16cm]{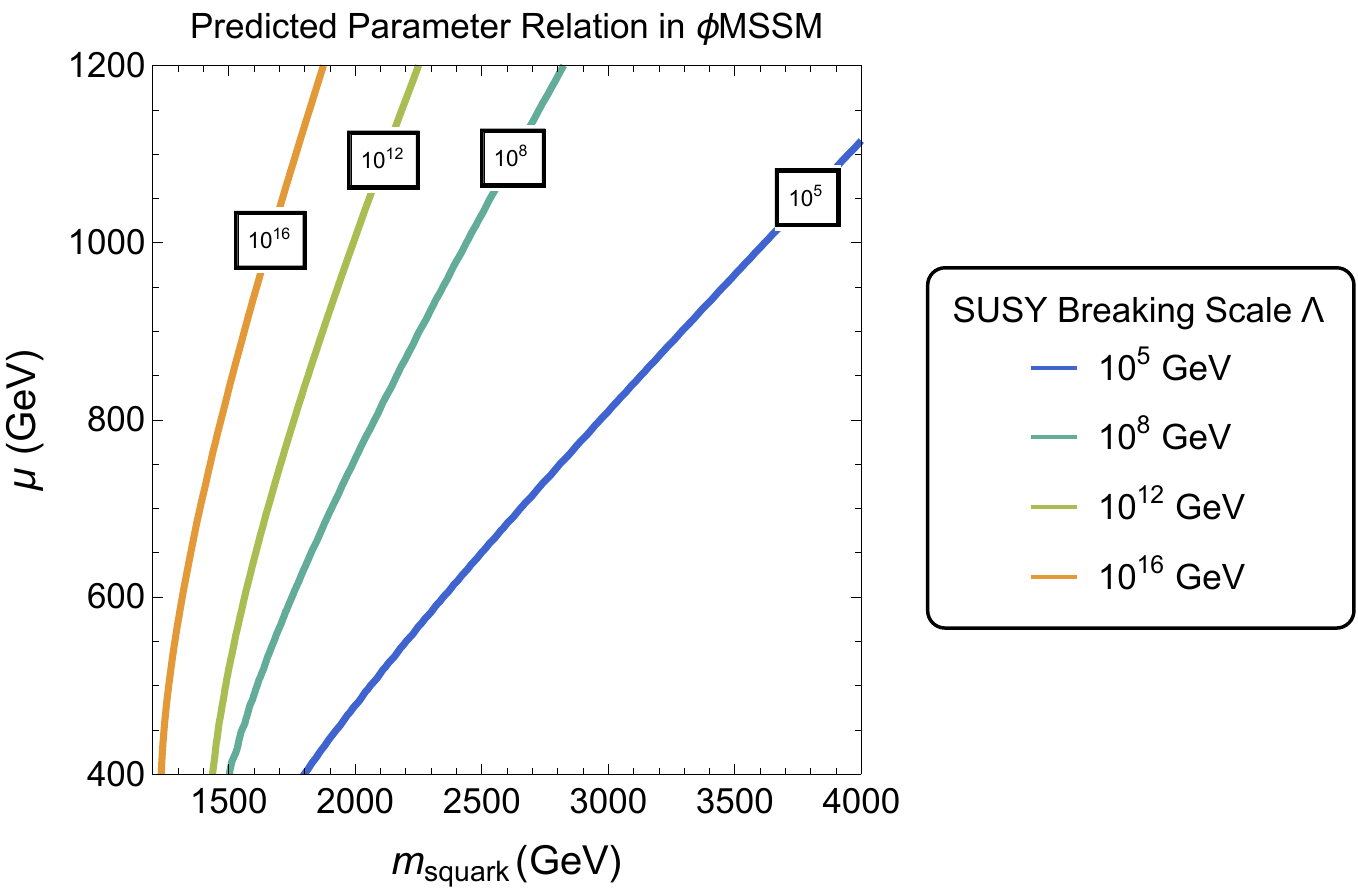}
 \end{center}
 \caption{Example of predictions for SUSY parameters in the $\phi$MSSM, from imposing the maximal vacuum energy condition. For simplicity we take all squark soft mass parameters to be equal, i.e.  $m_{\tilde{Q}_i}^2 = m_{\tilde{U}_i}^2 = m_{\tilde{D}_i}^2 \equiv m_{\rm squark}^2$ for all generations $i$, and plot this on the horizontal axis. The vertical axis corresponds to the effective Higgsino mass parameter, $\mu \equiv \mu(\phi_{\rm max})$. The contours indicate the allowed values for $\mu$ and $m_{\rm squark}$ for various values of the SUSY mediator scale $\Lambda$ in GeV. The other SUSY parameters are fixed at the low scale (1 TeV) as follows: $\tan \beta = 3$, $m_{A_0} = 600 \GeV$, $M_1 = M_2 = 1 \text{ TeV}$, $M_3 = 3\text{ TeV}$, $m_{\tilde{L}_i}^2 = m_{\tilde{E}_i}^2 = \left(1.5 \text{ TeV}\right)^2$, $a_i = 0$.}
 \label{fig:LambdaContours}
 \end{figure}
 
In \Fig{fig:LambdaContours}, we show an example of the predictions of the $\phi$MSSM for the SUSY spectrum by plotting contours of the possible values of the Higgsino mass parameter $\mu$ and the squark masses for various values of the SUSY mediation scale $\Lambda$, while keeping the weak scale, $\tan \beta$, the pseudoscalar Higgs mass $m_{A_0}$ and the other low-energy SUSY parameters like the gaugino soft masses fixed.\footnote{Here we do \emph{not} impose the condition that the lightest Higgs boson mass is 125 GeV. As is well known, a fully realistic SUSY model without very heavy stops etc. must involve some extension of the MSSM in order to achieve the observed Higgs mass, {\it e.g.} \Refs{Barbieri:2006bg,Batra:2003nj}.}

\begin{figure}
		  \centering
        \subfigure[]{
                \includegraphics[width=8cm]{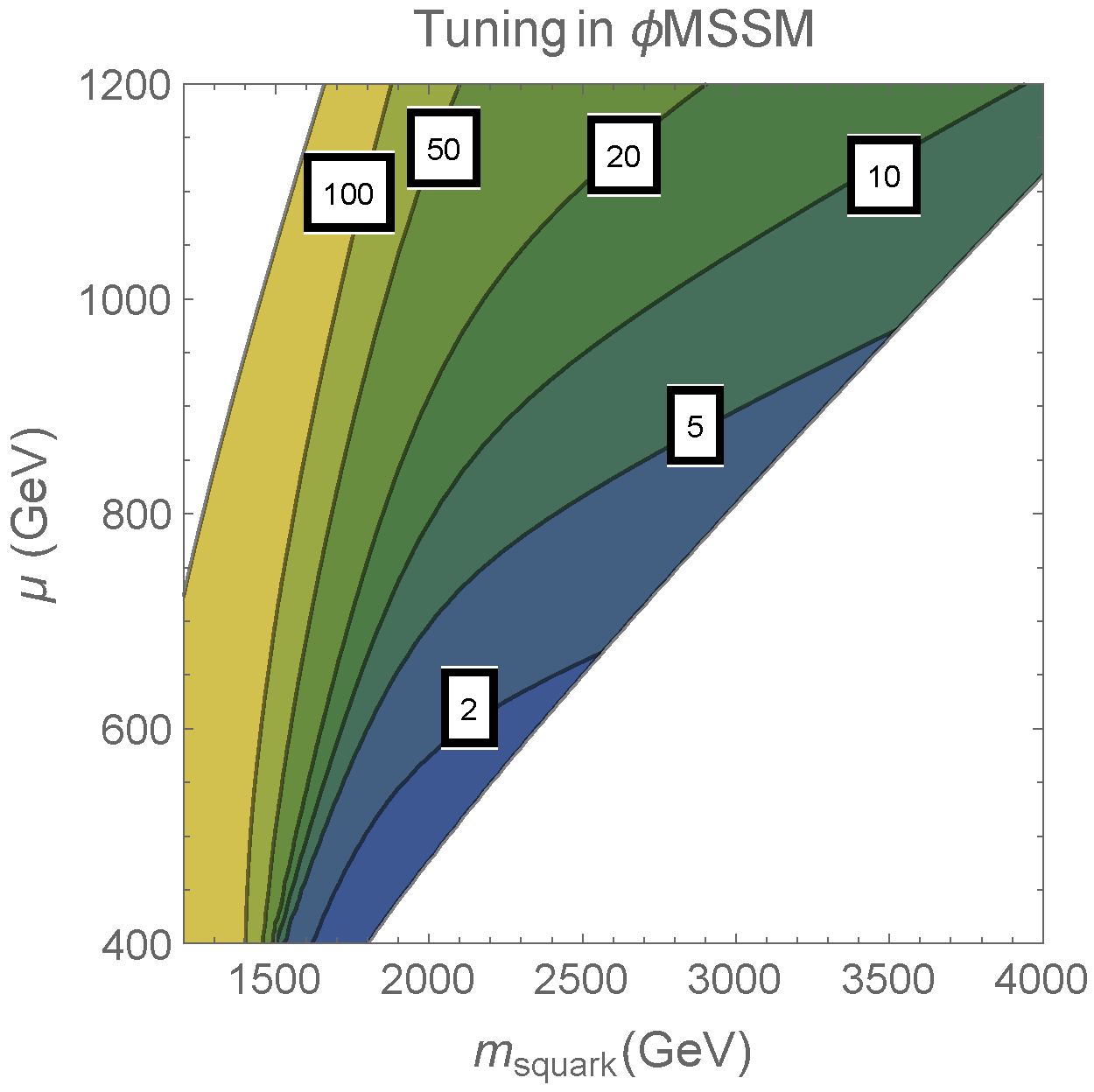}
                \label{fig:tuningA}
		  }
               \hspace{3mm} 
 	\subfigure[]{
                \includegraphics[width=8cm]{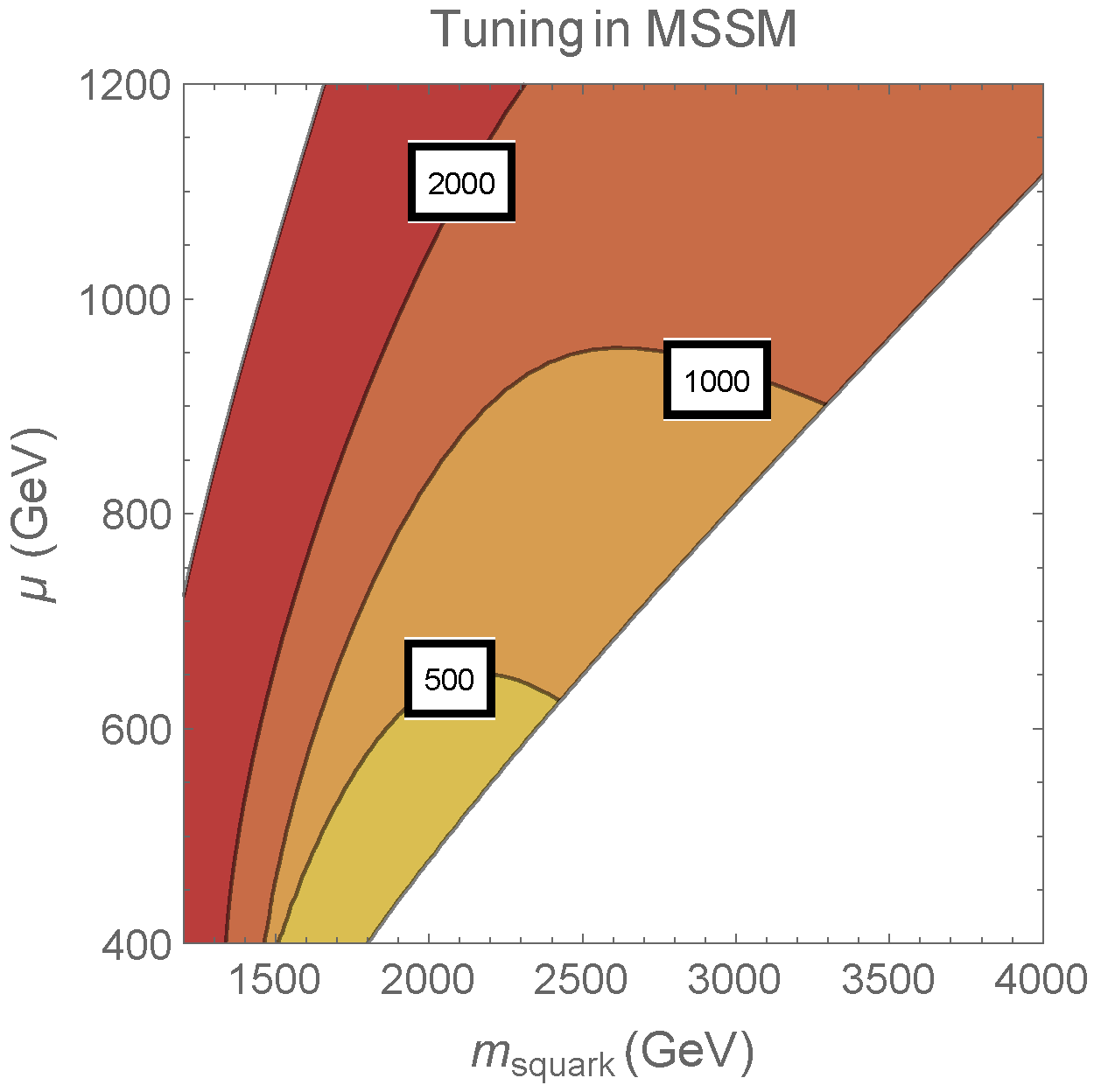}
                \label{fig:tuningB}
		  }
\caption{Degree of fine-tuning for the parameter space depicted in \Fig{fig:LambdaContours}, for both the $\phi$MSSM and the regular MSSM, {\it i.e.} with $\mu$ a fixed input parameter (right). The low-energy SUSY parameters are fixed as in Fig.~\ref{fig:LambdaContours}, with the SUSY mediator scale at each point set to the value predicted consistent with maximal vacuum energy. Only the region with $100\TeV < \Lambda< \mPl$ is shown. In the left plot the fine-tuning is computed by varying the soft parameters at the SUSY mediator scale while also varying $\mu$ so as to satisfy the maximal vacuum energy condition \Eq{eq:dVdmu}, while in the right plot $\mu$ is held fixed as the soft parameters are varied. The fine-tuning measure plotted here is $\sqrt{\Sigma_i \left( \frac{d \log m_Z^2}{d \log \theta_i} \right)^{2} }$, where the variables $\theta_i$ range over $m_{H_u}^2, m_{H_d}^2, b_0, m_Q^2, m_U^2, M_3$ (as evaluated at the high scale $\Lambda$). 
}        
\label{fig:tuning}
\end{figure}

Just as in the regular MSSM, one can compute the level of fine-tuning in the $\phi$MSSM, quantifying how much the prediction for the weak scale varies as the input parameters, {\it i.e.}~the SUSY parameters at the mediator scale $\Lambda$, are varied (\Fig{fig:tuning}). As claimed in \Sec{sec:hierarchy}, the SUSY masses can considerably heavier in this model compared to the MSSM without introducing fine-tuning. This can be understood as follows: suppose that some spectrum of SUSY particles is observed. This can be fit using the regular MSSM, but the resulting parameters may be tuned, in the sense that if an input parameter such as the high-scale value of $m_{H,u}^2$ were varied, then the predicted $Z$ mass would change by a much larger fractional amount. Now consider interpreting the same spectrum in terms of the $\phi$MSSM. The same input parameters are required; however if one imagines varying a parameter such as $m_{H,u}^2$, then the background value of $\phi$ will have to change such that \Eq{eq:dVdmu} remains satisfied. When $m_{H,u}^2$ is varied, the potential $\Vvq$ obtained from running \Eq{eq:betaV} will also vary, by some amount which includes a loop factor suppression (though with log-enhancement from running). So to preserve the relation $\Vvc'(\phi) + \Vvq'(\phi) = 0$, the slope of $\Vvc$ in \Eq{eq:VeffMSSM}, which is approximately proportional to $m_Z^2(\phi)$, should also vary by the same loop-suppressed amount. So despite the fact that varying $m_{H,u}^2$ at fixed value of $\phi$ would cause $m_Z^2$ to vary at tree-level, in the $\phi$MSSM the field $\phi$ will adjust when $m_{H,u}^2$ is varied so that $m_Z^2(\phi)$ varies by only a loop-suppressed amount.           

\Fig{fig:tuning} shows contours of fine-tuning in the same slice of parameter space as \Fig{fig:LambdaContours}. On the left we show the actual fine-tuning of the $\phi$MSSM model. On the right we show how fine-tuned the same spectrum would appear if interpreted within the regular MSSM, {\it i.e.} without allowing $\phi$ to adjust when the input parameters are varied. In all regions of parameter space the $\phi$MSSM model is much less tuned than the MSSM interpretation of the same spectrum. In particular, for the parameters shown the $\phi$MSSM is essentially ``untuned'' when the mediation scale is relatively low (compare to \Fig{fig:LambdaContours}), since in this case there is less RG running and $\Vvq$ is smaller. 

In addition to addressing the hierarchy, the $\phi$MSSM can reproduce the other potential advantages of TeV-scale SUSY. Gauge coupling unification for example proceeds just as in the MSSM, since the new $\phi$ field is a singlet with small couplings. The prospects for a dark matter candidate improve: a pure Higgsino LSP can produce the observed DM abundance for mass $\approx 1 \TeV$~\cite{Cirelli:2007xd}, which requires fine-tuning within the MSSM but can be relatively natural in the $\phi$MSSM (as in \Fig{fig:tuningA}).

\subsection{Collider Physics}
\label{sec:collider}

The hypothesis of maximal vacuum energy has significant implications for collider searches for SUSY. Traditional naturalness, which requires the Higgsinos to be not much above the weak scale and the stops to be lighter than $\sim \text{TeV}$, is strained by the strong bounds on squark production from the LHC, see {\it e.g.} \Refs{Sirunyan:2017cwe, Aaboud:2017vwy}. This has motivated searches for so-called ``natural SUSY'' spectra~\cite{Brust:2011tb, Papucci:2011wy}, in which the stops are much lighter than the squarks of the first two generations, but even this is highly constrained by searches for top-rich final states, for example \Ref{Sirunyan:2017xse, Aaboud:2017ayj, Sirunyan:2017pjw, Aaboud:2017hrg}. With the maximal vacuum energy hypothesis however, the Higgsino mass $\mu$, the Higgs soft parameters and the stop masses can all be roughly a loop factor higher than the natural MSSM range without fine-tuning of parameters (as in \Fig{fig:tuningA}). Searches for SUSY at much higher mass scales are therefore still motivated by naturalness in this context, and the possibilities for allowed ``natural'' spectra are broader. For example, instead of requiring the stops to be much lighter than the other squarks, one could have all the squarks approximately degenerate at mass scale of a few TeV without incurring extreme tuning (as depicted in \Fig{fig:tuningA}), so naturalness no longer directly implies top-rich final states. One class of searches that become particularly motivated are those targeting ``compressed'' spectra, in which the LSP is relatively close in mass to the original pair-produced superpartners so that the visible energy in the event is suppressed. Traditional MSSM naturalness requires the Higgsino to be close to the weak scale, so that the LSP can be no heavier than that; this condition is relaxed in the extremal vacuum energy context. Compressing the spectrum often greatly weakens the bounds on superpartner masses (see {\it e.g.} \Refs{Sirunyan:2017cwe, Aaboud:2017vwy}), which could even further reduce the tension between SUSY naturalness and current LHC null results.    

The extremal vacuum energy condition eq.~\ref{eq:dVdmu} implies a single ``sum rule'' constraint on the SUSY parameters, including the SUSY mediation scale $\Lambda$ (see {\it e.g.} \Fig{fig:LambdaContours}). Directly testing this constraint will require knowledge of the SUSY mediation scale which is likely out of reach of forseeable colliders (though certain observables such as the width of a long-lived NLSP could provide hints). Instead, given the low-energy SUSY parameters, one can predict the mediator scale assuming the maximal vacuum energy hypothesis, as in \Fig{fig:LambdaContours}.  One can then determine whether the spectrum in question is fine-tuned or not under this hypothesis (Figure~\ref{fig:tuningA}). A SUSY spectrum which is finely tuned in terms of the standard MSSM (Figure~\ref{fig:tuningB}) but not tuned under the maximal vacuum energy condition would be suggestive evidence for the latter hypothesis. Assuming no large hierarchies in the SUSY breaking masses, the most relevant parameters for making this determination are the Higgsino mass, the Higgs soft masses, and the stop and gluino masses. 

\section{Conclusions}
\label{sec:conclusion}

We have argued in this work that the observed value of the weak scale could potentially be explained by a ``principle of maximum vacuum energy.'' The argument involves two independent points. The first is that if the Higgs sector parameters are scanned by some background field $\phi$, it is generic for the vacuum energy as a function of this field to be maximized at a low value of the weak scale. This was explored in generality in \Sec{sec:crit} and for a specific MSSM-like model in \Sec{sec:SUSY}. The second necessary ingredient for our argument is some mechanism to place the field at or near this local maximum of vacuum energy. We proposed several distinct mechanisms to realize this in \Sec{sec:mech}. Most of these rely on the effect of the vacuum energy on cosmological evolution, {\it i.e.} they make use of the fact that gravity is sensitive to the value of the vacuum energy and not just its gradients.

As in models of cosmological relaxation~\cite{Graham:2015cka} or the model of \Ref{Geller:2018xvz}, the mechanisms we have discussed for placing the Universe in a vacuum with low weak scale act very early in cosmological history (before or during inflation), and are thus not easily tested by experiment, though some potential direct probes of the field $\phi$ were discussed in \Sec{sec:pheno}. However, independent of the underlying mechanism, the assumption that the vacuum energy is maximized with respect to $\phi$ gives testable predictions for TeV-scale physics, in the context of any model which predicts the weak scale in terms of other observable parameters. In our approach the location of the maximum of vacuum energy is determined by the radiative corrections to the potential of $\phi$. These will arise at two-loop from heavy states coupled to the Higgs ({\it e.g.}~the top squark), or at one-loop from Higgs ``partners'' ({\it e.g.}~the Higgsino). In any particular model the masses of such states must satisfy some additional ``sum rule'' constraint in order to maximize the vacuum energy. This constraint on the parameter space, which is responsible for setting the weak scale to a low value, is experimentally testable and falsifiable. 

We have explored this constraint more quantitatively for a particular model in which $\phi$ scans parameters of the MSSM (see {\it e.g.} \Fig{fig:LambdaContours}). More generally however one could consider scanning parameters within different models addressing the hierarchy problem, such as composite Higgs models. As we have emphasized, our argument that the vacuum energy as a function of $\phi$ can have a maximum for low weak scale applies for quite general choice of Higgs sector and parameters which scan, provided that the potential $\Vv(\phi)$ is generated due to the couplings of $\phi$ to the Higgs sector. Thus we expect that our approach can generically allow for higher scales for new physics in other models which address the hierarchy problem, without requiring fine-tuning. It would be interesting to explicitly realize the maximal vacuum energy approach in models besides the MSSM and explore the resulting predictions.   

As discussed in \Sec{sec:hierarchy}, in the absence of fine-tuning, the expected mass scale of new physics is enhanced by at most a square root loop factor ($\sim 4 \pi$) when the maximal vacuum energy constraint is imposed. This shifts the ``target'' for collider searches for new physics associated with the hierarchy problem towards the upper end of the LHC's energy reach, or even into a range requiring a next-generation $\sim 100 \TeV$ collider. Discovering top partners etc. at these energy scales would usually suggest that the weak scale is tuned. However, if the ``principle of maximum vacuum energy'' is at work, then with the right model interpretation including a scanning $\phi$ one could experimentally verify that the constraint of maximal vacuum energy is satisfied, and that the model is not actually tuned when this constraint is imposed (as in \Fig{fig:tuning}). This would be highly suggestive that some mechanism is indeed acting to maximize the vacuum energy with respect to $\phi$, revealing a new connection between the physics of the weak scale and cosmology in a scalar field landscape.

\begin{center}
{\bf Acknowledgements}
\end{center}
The authors thank Nathaniel Craig, Michael Geller, Eric Kuflik, Grant Remmen and Kathryn Zurek for helpful discussions. CC is supported by a Sloan Research Fellowship and a DOE Early Career Award under Grant No. DE-SC0010255. PS is supported by the DuBridge Fellowship of the Walter Burke Institute for Theoretical Physics. This material is based upon work supported by the U.S. Department of Energy, Office of Science, Office of High Energy Physics, under Award Number DE-SC0011632.  
\appendix

\section{Solutions of the Modified Fokker-Planck Equation}
\label{sec:appendix}

In this appendix we consider solutions to the modified Fokker-Planck equation
\eq{
\frac{\partial P}{\partial t} = \frac{\partial}{\partial \phi} \left[ \frac{H(\phi)^3}{8\pi^2} \frac{\partial P}{\partial \phi} + \frac{V'(\phi)}{3 H(\phi)} P \right] + 3 H(\phi) P.
\label{eq:FP2}
}
We will first give the nearly-exact solution to this equation for a quadratic potential $V(\phi) = m^2 \phi^2/2$, and then discuss its relevance to the various scenarios described in \Sec{sec:inflation}. We will allow either sign for $m^2$ in this analysis. For $m^2 > 0$, a stable solution centered at $\phi = 0$ will exist if classical rolling dominates over faster expansion of regions higher on the potential. For $m^2 < 0$ the solution is always decaying, but remains peaked at $\phi = 0$ due to the faster expansion at that point. With these results for quadratic potentials we can describe both the effect of the EWSB vacuum energy $\sim - g^2 \phi^2/\lambdaH$ as well the periodic bare potential $M^4 \cos \phi/f$ in certain limits. 

Our analysis essentially follows that of Appendix B of \Ref{Graham:2018jyp} (which in turn draws on the approach of \Ref{StopyraThesis} for solving the ordinary Fokker-Planck equation). As in \Sec{sec:inflation}, the Hubble parameter can be expanded as $H(\phi) \equiv \Hinf + \frac{1}{2} \frac{V(\phi)}{\Vinf} \Hinf = \Hinf + \frac{1}{6} \frac{V(\phi)}{\Hinf \mPl^2} $. Let us absorb the $\phi$-independent spacetime expansion due to $\Hinf$ by a redefinition $\tilde{P}(\phi, t) \equiv P(\phi, t) \exp(3 \Hinf t)$. Furthermore, we will ignore the $\phi$ dependence of $H$ in the diffusion and drift terms of the Fokker-Planck equation, which is a good approximation since its fractional variation is small. Then the modified Fokker-Planck equation becomes
\eq{
\frac{\partial \tilde{P}}{\partial t} = \frac{\Hinf^3}{8\pi^2} \frac{\partial^2 \tilde{P}}{\partial \phi^2} + \frac{1}{3 \Hinf} \frac{\partial}{\partial \phi} \left[ V'(\phi) \tilde{P} \right] + \frac{1}{2} \frac{V(\phi)}{\Hinf \mPl^2} \tilde{P}
}
As shown in \Ref{Graham:2018jyp}, we can now make a field redefinition $\tilde{P}(\phi, t) \equiv \exp[-\nu(\phi)]\psi(\phi, t)$ where $\nu(\phi) \equiv 4\pi^2 V(\phi)/3\Hinf^4$ to obtain
\eq{
-\frac{4\pi^2}{\Hinf^3} \frac{\partial \psi}{\partial t} = -\frac{1}{2} \frac{\partial^2 \psi}{\partial \phi^2} + \frac{1}{2} \left[-\nu''(\phi) +\nu'(\phi)^2 -\frac{3}{\mPl^2} \nu(\phi) \right] \psi
\label{eq:Schr}
}
which is a Wick-rotated version of the Schr\"odinger equation. In particular, for $V(\phi), \nu(\phi) \propto \phi^2$, it resembles the Schr\"odinger equation for a harmonic oscillator.  We can consider an eigenvalue decomposition of the solutions, {\it i.e.} $\psi(\phi, t) \equiv \sum_n c_n \psi_n(\phi) e^{- \Gamma_n t}$ where the eigenvalues $\Gamma_n$ give decay rates instead of frequencies. Then for $V(\phi) = \frac{1}{2} m^2 \phi^2$, the $\psi_n$ satisfy the time-independent Schr\"odinger equation for a harmonic oscillator:
\eq{
\frac{4\pi^2}{\Hinf^3} \Gamma_n \psi_n = -\frac{1}{2} \frac{\partial^2 \psi_n}{\partial \phi^2} + \left[-\frac{2 \pi^2 m^2}{3 \Hinf^4} + \left(\frac{8\pi^4 m^4}{9 \Hinf^8} - \frac{\pi^2 m^2}{\Hinf^4 \mPl^2} \right)\phi^2 \right] \psi_n
}
We can put this into a particularly familiar form with some redefinitions:
\eq{
E_n \psi_n = -\frac{1}{2} \frac{\partial^2 \psi_n}{\partial \phi^2} + \frac{1}{2} \omega^2 \phi^2
}
\eq{
\omega^2 \equiv \left( \frac{4 \pi^2 m^2}{3 \Hinf^4} \right)^2 \left(1 - \delta \right) \qquad \delta \equiv \frac{9 \Hinf^4}{8 \pi^2 m^2 \mPl^2} \qquad E_n \equiv \frac{2 \pi^2 m^2}{3 \Hinf^4} + \frac{4\pi^2}{\Hinf^3} \Gamma_n 
}

For $\omega^2 > 0$ this has solutions localized around $\phi = 0$, while $\omega^2 < 0$ indicates that $\phi = 0$ is not an attractor point for $P(\phi, t)$. Note that both possibilities can occur for $m^2 > 0$, depending on whether classical rolling or differential expansion dominates; the latter occurs if $\delta > 1$. For $m^2 < 0$ we always have $\omega^2 > 0$. 

For $\omega^2 > 0$ the eigenvalues are the familiar $E_n = (n+1/2)\omega$, giving decay rates
\eq{
\Gamma_n = - \frac{m^2}{6 \Hinf} + \frac{\Hinf^3}{4\pi^2} \omega (n+ 1/2)
}
The late time behavior for $\psi(\phi, t)$ is dominated by the lowest eigenmode $n=0$ with the lowest decay rate, which has the same Gaussian profile as the ground state of the harmonic oscillator. Changing back to $P(\phi,t)$, this steady-state solution is
\begin{gather}
P(\phi, t \rightarrow \infty) \propto \exp(3 \Hinf t) \exp(- \Gamma_0 t) \exp \left[ -\phi^2/2\sigma_\phi^2 \right] \\
\sigma_\phi^2 = \left(\omega + \frac{4 \pi^2 m^2}{3 \Hinf^4} \right)^{-1} = \frac{3\Hinf^4}{4\pi^2} \left(|m|^2\sqrt{1-\delta}  + m^2 \right)^{-1}
\end{gather}
Note that $\sigma_\phi^2$ is always positive for either sign of $m^2$ (so long as $\omega^2$ is positive). 

Since the decay rate eigenvalues for the quadratic potential have constant spacing $\Delta \Gamma =\frac{\Hinf^3}{4\pi^2} \omega$, the typical relaxation time for a generic initial $\psi(\phi,t)$ to approach the steady-state solution is $\tau_r \sim \Delta \Gamma^{-1} = \frac{4 \pi^2}{\Hinf^3} \omega^{-1}$. We can write this as 
\eq{
\tau_r \sim \frac{8 \pi^2}{3} \frac{\mPl^2}{\Hinf^3} \frac{|\delta|}{\sqrt{1-\delta}}.
}
Recall however that non-eternal slow-roll inflation can typically last for a maximum of $\mPl^2/\Hinf^2$ e-folds (\Eq{eq:NeNonEternal}), {\it i.e.} a maximum time $t_{\rm max} \sim \mPl^2/\Hinf^3$. If we want to approach the steady-state solution within such a period of inflation, we must have $\tau_r \ll t_{\rm max}$, or $|\delta| \ll 1$. We will focus on this case in what follows. We then have $\tau_r \sim \Hinf/m^2$ (though as discussed below, for a more realistic $V(\phi)$ some initial distributions will take longer to relax to the steady state).

Now let us consider the various limits of the above for both $m^2 > 0$ and $m^2 < 0$:
\begin{itemize}
\item $m^2 > 0$: For $\delta < 1$, $P(\phi, t)$ approaches a steady-state distribution with width $\sigma_\phi \lesssim \Hinf^2/m$ around the minimum of the potential. We have $\Gamma_0 \approx 0$ for small $\delta$ as probability is conserved in the $\mPl \rightarrow \infty$ limit. For $\delta > 1$ ($m \lesssim \Hinf^2/\mPl$) however, $\omega$ becomes negative indicating that $P(\phi,t)$ is instead driven up the potential away from $\phi = 0$. 

\item $m^2 < 0$: For $|\delta| \ll 1$, the steady-state has width $\sigma_\phi \approx 2 \mPl/\sqrt{3}$, independent of $m$ or $\Hinf$. For $|\delta| \gg 1$, we have $\sigma_\phi \sim \sqrt{\mPl/m} \Hinf$, which is $\gg \mPl$ for $\delta \gg 1$. 

\end{itemize}  

A general (smooth) potential $V(\phi)$ can be approximated by a quadratic in some neighborhood of an extremum. The steady-state solutions discussed above are valid if the quadratic form remains a good approximation over the width $\sigma_\phi$. 

However, the time required for an arbitrary initial distribution to approach this steady-state can depend on the global shape of the potential.\footnote{We thank the authors of \Ref{Geller:2018xvz}, in particular Michael Geller, for bringing this to our attention and for the essential points of the following argument.} For example, suppose that $V(\phi)$ has a negative quadratic form only up to some value $\sim \phi_{\rm max}$ where a local minimum or ``valley'' forms. Then the difference in expansion rate between $\phi = 0$ and points lower on the potential is at most $\Delta H \sim m^2 \phi_{\rm max}^2/\Hinf\mPl^2$, rather than being unbounded as for a pure quadratic potential. Suppose that the initial distribution for $\phi$ is centered at some $O(1)$ distance towards the valley with some width $\sigma_0$, e.g. $P(\phi,t=0) \sim \exp [-(\phi-\phi_{\rm max})^2/\sigma_0^2]$. Then in the limit of ignoring diffusion we can bound the time $t$ required for the region near $\phi = 0$ to dominate by requiring the maximal differential expansion factor $e^{3 \Delta H t}$ to win out over the initial suppression $e^{-\phi_{\rm max}^2/\sigma_0^2}$, giving $t \gtrsim (\Hinf/m^2)(\mPl^2/\sigma_0^2)$. Thus the above estimate of the time to approach the steady state, $\sim H/m^2$, actually only holds for $\sigma_0 \gtrsim \mPl$ (though given this, we can take $\phi_{\rm max} \gg \mPl$). If the initial $P(\phi, t=0)$ is very narrow, e.g. a delta function, then it can diffuse out to width $\mPl$ after time $\sim \mPl^2/\Hinf^3$, but this is comparable to the maximum duration of slow-roll inflation. In the Schr\"odinger equation approach, this breakdown of the na\"ive relaxation time estimate occurs because the effective potential of \Eq{eq:Schr} turns over when the physical potential $V(\phi)$ deviates from a quadratic, creating a finite energy barrier rather than an infinite harmonic well. This allows for eigenmodes localized away from $\phi = 0$, which could have large overlap with the initial distribution.

For this work we are also interested in how a perturbing periodic potential $M^4 \cos \phi/f$ (\Sec{sec:minima}) affects the solutions of the modified F-P equation. \Ref{Graham:2018jyp} explicitly considers solutions to the F-P equation with a cosine potential; here we will reproduce their parametric results by taking appropriate limits of the above results for a quadratic potential. The regions near minima or maxima of the $M^4 \cos \phi/f$ potential can be approximated as quadratics with mass $m^2 = \pm M^4/f^2$ respectively, though this breaks down of course for field ranges $\gtrsim f$. We then have $|\delta| \sim \frac{\Hinf^4}{M^4} \frac{f^2}{\mPl^2}$. Let us consider two parametric regimes that are of interest to us:
\begin{itemize}

\item $M^4 \gg \Hinf^4$, $f \ll \mPl$: In this case $|\delta| \ll 1$. The above solution for $m^2 >0, \delta \ll 1$ will be a good approximation near the minima since $\sigma_\phi \sim \left(\Hinf/M\right)^2 f \ll f$. The solution for $m^2 < 0$ will not be valid near the maxima however since in that case $\sigma_\phi \sim \mPl \gg f$. So we expect that $P(\phi,t)$ will accumulate around the minima of the potential, but not the maxima.

\item  $M^4 \ll \Hinf^4$, $f \ll \mPl$: In this case the quadratic approximations give $\sigma_\phi \gg f$ for both attractor solutions, so they are not valid around either the minima or maxima. So the potential is essentially negligible and $P(\phi,t)$ diffuses uniformly. 

\end{itemize}  
Therefore, so long as $f \ll \mPl$, $P(\phi,t)$ will not localize around maxima of the cosine potential. 

\section{Comparison to Other Models}
\label{sec:compare}

Our approach of determining the weak scale as a point of extremal vacuum energy shares some features with both the idea of cosmological relaxation~\cite{Graham:2015cka} as well as the recent work of Geller, Hochberg and Kuflik~\cite{Geller:2018xvz}. However, the actual principles determining the weak scale are quite different in these models. In this section we will briefly review and compare these three approaches.

\begin{figure}
\begin{center}
  \includegraphics[width=17cm]{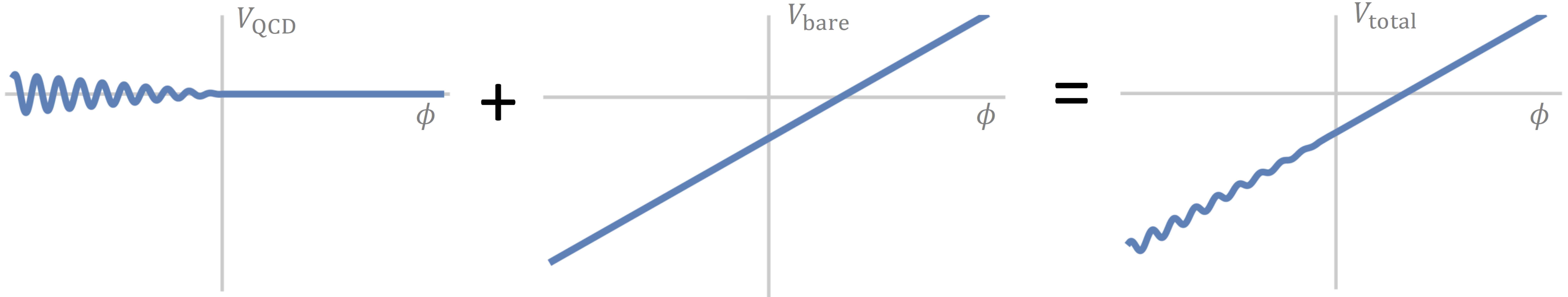}
 \end{center}
 \caption{Illustrative sketch of the contributions to the $\phi$ potential in the relaxion model~\cite{Graham:2015cka}. A QCD-axion-like potential (left) turns on in the region where the Higgs mass becomes negative. After including some bare $H$-independent potential (middle), the total vacuum energy can have weak-scale minima as an attractor region.   
}
 \label{fig:relaxion}
 \end{figure}

In all three models, the Higgs mass is assumed to be scanned linearly by some light field $\phi$. In the ``relaxion'' model~\cite{Graham:2015cka}, $\phi$ also has a QCD-axion-like coupling, which gives an effective potential of $\sim f_\pi^2 m_\pi^2(H) \cos \frac{\phi}{f}$ when EW symmetry is broken (where $m_\pi^2(H)$ is the pion mass, which depends on the Higgs vev). The behavior of the vacuum energy in the relaxion model is illustrated in \Fig{fig:relaxion}. The QCD-axion-like potential (leftmost plot) turns on when the Higgs mass turns negative, creating local minima. When a bare ($H$-independent) linear potential for $\phi$ is also included (middle plot), the local minima only start appearing at some finite value of the weak scale (rightmost plot). Given a long period of cosmic inflation, the weak-scale minima can be cosmological attractors as $\phi$ will slow-roll to them starting from any point higher up the potential.

The key ingredient distinguishing relaxion models from the approach we have taken is the highly nonlinear, oscillatory response of the vacuum energy when the Higgs vev turns on. This occurs because the QCD axion coupling $\propto \cos \phi/f$ violates our assumption in section \Sec{sec:crit} that the scanning of the Higgs sector parameters by $\phi$ can be linearized, so  the result in \Eq{eq:Vpp} indicating that the vacuum energy is concave down does not apply. Note that the periodic and non-periodic potentials for $\phi$ play opposite roles in our model compared to the relaxion. In our work we invoke an oscillatory \emph{bare} potential for $\phi$ in order to create local minima (\Sec{sec:minima}); however this potential is $H$-independent, and the value of the weak scale is determined by the position of the maximum in vacuum energy that develops in response to EWSB (\Fig{fig:plateau}). In contrast, relaxion models must \emph{avoid} any bare, $H$-independent periodic potential for $\phi$, as this could stop the relaxion's rolling before it reaches the region where EWSB is broken.

The model of \Ref{Geller:2018xvz} is more similar to our approach in that it arranges for a maximum of the vacuum energy to arise at low weak scale due to a phase transition. The phase transition invoked in \Ref{Geller:2018xvz} is not however electroweak symmetry breaking, but rather a first-order phase transition in some new scalar $a$. The field $a$ is assumed to have a QCD-axion-like coupling giving a $H$-dependent potential $\sim f_\pi^2 m_\pi^2(H) \cos \frac{a}{f}$, plus some additional bare potential, just as in the relaxion. \Ref{Geller:2018xvz} assumes that the intial values of $\phi$ and $a$ are such that the weak scale is large and $a$ is trapped in a metastable vacuum created by QCD-axion-like potential. As the weak scale scans to lower values (through random-walking of $\phi$ during inflation), this potential weakens until it no longer traps $a$, causing a first-order phase transition when $a$ becomes free to roll to its  global minimum. The effective shape of the vacuum energy in this model is sketched in \Fig{fig:relaxion}. The leftmost plot shows the discontinuous drop in energy as the weak scale is lowered due to the first-order phase transition. Adding some bare potential for $\phi$ as in the middle plot, the total vacuum energy can have a maximum at the point of the $a$ phase transition (rightmost plot), which can be arranged to occur for low weak scale.

 \begin{figure}
\begin{center}
  \includegraphics[width=17cm]{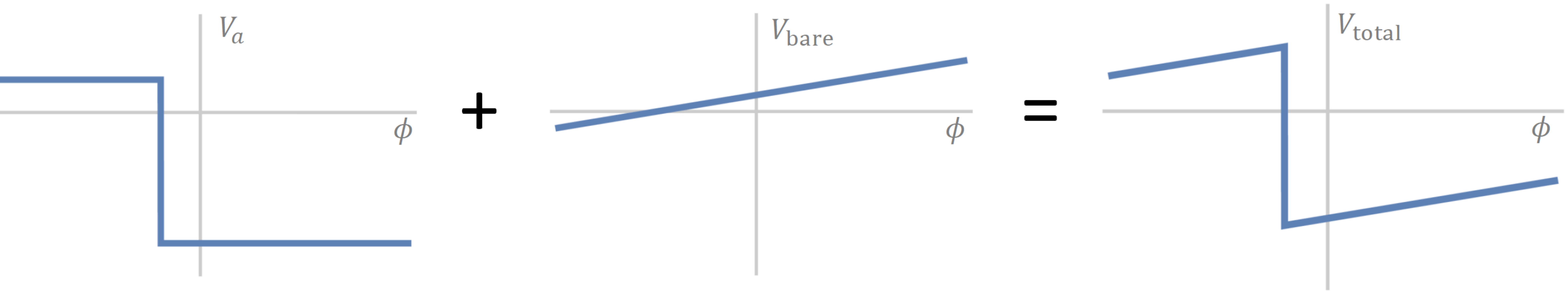}
 \end{center}
 \caption{Illustrative sketch of the contributions to the $\phi$ potential in the model of Geller, Hochberg and Kuflik~\cite{Geller:2018xvz}. There is a discontinuous drop in the vacuum energy as the weak scale is lowered due to a first-order phase transition in the field $a$. With the addition of a bare $\phi$ potential, the point of the phase transition becomes a maximum of vacuum energy. 
}
 \label{fig:GHK}
 \end{figure}
 
Both \Ref{Graham:2015cka} and \Ref{Geller:2018xvz} make use of the QCD axion potential, $\sim f_\pi^2 m_\pi^2(H) \cos \frac{\phi}{f}$, which vanishes when EW symmetry is unbroken. It is technically natural to have no bare, $H$-independent coefficient for $\cos \frac{\phi}{f}$ in the potential, since in the SM a chiral symmetry is restored when EW symmetry is unbroken. However, in both \Ref{Graham:2015cka} and \Ref{Geller:2018xvz}, the fields with axion-like couplings scan the strong CP angle $\theta$ like the usual QCD axion, but the additional non-periodic terms in the potential spoil the usual prediction that $\theta = 0$ at the potential minima. Instead the first minima correspond to $\theta \approx \pm \pi/2$. To avoid this prediction for the strong CP angle, other variations of the relaxion framework invoke other BSM physics to produce a coupling of $\cos \frac{\phi}{f}$ to $H$. In order to avoid generating a large $H$-independent $\cos \phi/f$ potential this new physics generically cannot be very heavy; for example in the ``non-QCD'' model of \Ref{Graham:2015cka}, the new states are a sector of vectorlike leptons with large couplings to the Higgs and masses less than a TeV. This may be compared to the parametrics for the scale of BSM physics in our approach as outlined in \Sec{sec:hierarchy} or discussed in a specific SUSY context in \Sec{sec:SUSY}. The actual role of TeV-mass BSM states is of course very different in our framework, where their spectrum directly determines the value of the weak scale.

\bibliographystyle{JHEP}

\bibliography{hierarchy_bib}

\providecommand{\href}[2]{#2}\begingroup\raggedright\begin{thebibliography}{10}

\bibitem{Vafa:1984xg}
C.~Vafa and E.~Witten, \emph{{Parity Conservation in QCD}},
  \href{https://doi.org/10.1103/PhysRevLett.53.535}{\emph{Phys. Rev. Lett.}
  {\bfseries 53} (1984) 535}.

\bibitem{Geller:2018xvz}
M.~Geller, Y.~Hochberg and E.~Kuflik, \emph{{Inflating to the Weak Scale}},
  \href{https://arxiv.org/abs/1809.07338}{{\ttfamily 1809.07338}}.

\bibitem{Kaloper:2011jz}
N.~Kaloper, A.~Lawrence and L.~Sorbo, \emph{{An Ignoble Approach to Large Field
  Inflation}}, \href{https://doi.org/10.1088/1475-7516/2011/03/023}{\emph{JCAP}
  {\bfseries 1103} (2011) 023}
  [\href{https://arxiv.org/abs/1101.0026}{{\ttfamily 1101.0026}}].

\bibitem{Dubovsky:2011tu}
S.~Dubovsky, A.~Lawrence and M.~M. Roberts, \emph{{Axion monodromy in a model
  of holographic gluodynamics}},
  \href{https://doi.org/10.1007/JHEP02(2012)053}{\emph{JHEP} {\bfseries 02}
  (2012) 053} [\href{https://arxiv.org/abs/1105.3740}{{\ttfamily 1105.3740}}].

\bibitem{Yonekura:2014oja}
K.~Yonekura, \emph{{Notes on natural inflation}},
  \href{https://doi.org/10.1088/1475-7516/2014/10/054}{\emph{JCAP} {\bfseries
  1410} (2014) 054} [\href{https://arxiv.org/abs/1405.0734}{{\ttfamily
  1405.0734}}].

\bibitem{Furuuchi:2015foh}
K.~Furuuchi, \emph{{Excursions through KK modes}},
  \href{https://doi.org/10.1088/1475-7516/2016/07/008}{\emph{JCAP} {\bfseries
  1607} (2016) 008} [\href{https://arxiv.org/abs/1512.04684}{{\ttfamily
  1512.04684}}].

\bibitem{Silverstein:2008sg}
E.~Silverstein and A.~Westphal, \emph{{Monodromy in the CMB: Gravity Waves and
  String Inflation}},
  \href{https://doi.org/10.1103/PhysRevD.78.106003}{\emph{Phys. Rev.}
  {\bfseries D78} (2008) 106003}
  [\href{https://arxiv.org/abs/0803.3085}{{\ttfamily 0803.3085}}].

\bibitem{McAllister:2008hb}
L.~McAllister, E.~Silverstein and A.~Westphal, \emph{{Gravity Waves and Linear
  Inflation from Axion Monodromy}},
  \href{https://doi.org/10.1103/PhysRevD.82.046003}{\emph{Phys. Rev.}
  {\bfseries D82} (2010) 046003}
  [\href{https://arxiv.org/abs/0808.0706}{{\ttfamily 0808.0706}}].

\bibitem{Flauger:2009ab}
R.~Flauger, L.~McAllister, E.~Pajer, A.~Westphal and G.~Xu, \emph{{Oscillations
  in the CMB from Axion Monodromy Inflation}},
  \href{https://doi.org/10.1088/1475-7516/2010/06/009}{\emph{JCAP} {\bfseries
  1006} (2010) 009} [\href{https://arxiv.org/abs/0907.2916}{{\ttfamily
  0907.2916}}].

\bibitem{Choi:2014rja}
K.~Choi, H.~Kim and S.~Yun, \emph{{Natural inflation with multiple
  sub-Planckian axions}},
  \href{https://doi.org/10.1103/PhysRevD.90.023545}{\emph{Phys. Rev.}
  {\bfseries D90} (2014) 023545}
  [\href{https://arxiv.org/abs/1404.6209}{{\ttfamily 1404.6209}}].

\bibitem{Higaki:2014pja}
T.~Higaki and F.~Takahashi, \emph{{Natural and Multi-Natural Inflation in Axion
  Landscape}}, \href{https://doi.org/10.1007/JHEP07(2014)074}{\emph{JHEP}
  {\bfseries 07} (2014) 074} [\href{https://arxiv.org/abs/1404.6923}{{\ttfamily
  1404.6923}}].

\bibitem{Kaplan:2015fuy}
D.~E. Kaplan and R.~Rattazzi, \emph{{Large field excursions and approximate
  discrete symmetries from a clockwork axion}},
  \href{https://doi.org/10.1103/PhysRevD.93.085007}{\emph{Phys. Rev.}
  {\bfseries D93} (2016) 085007}
  [\href{https://arxiv.org/abs/1511.01827}{{\ttfamily 1511.01827}}].

\bibitem{Graham:2015cka}
P.~W. Graham, D.~E. Kaplan and S.~Rajendran, \emph{{Cosmological Relaxation of
  the Electroweak Scale}},
  \href{https://doi.org/10.1103/PhysRevLett.115.221801}{\emph{Phys. Rev. Lett.}
  {\bfseries 115} (2015) 221801}
  [\href{https://arxiv.org/abs/1504.07551}{{\ttfamily 1504.07551}}].

\bibitem{OConnell:2006rsp}
D.~O'Connell, M.~J. Ramsey-Musolf and M.~B. Wise, \emph{{Minimal Extension of
  the Standard Model Scalar Sector}},
  \href{https://doi.org/10.1103/PhysRevD.75.037701}{\emph{Phys. Rev.}
  {\bfseries D75} (2007) 037701}
  [\href{https://arxiv.org/abs/hep-ph/0611014}{{\ttfamily hep-ph/0611014}}].

\bibitem{Batell:2009jf}
B.~Batell, M.~Pospelov and A.~Ritz, \emph{{Multi-lepton Signatures of a Hidden
  Sector in Rare B Decays}},
  \href{https://doi.org/10.1103/PhysRevD.83.054005}{\emph{Phys. Rev.}
  {\bfseries D83} (2011) 054005}
  [\href{https://arxiv.org/abs/0911.4938}{{\ttfamily 0911.4938}}].

\bibitem{Piazza:2010ye}
F.~Piazza and M.~Pospelov, \emph{{Sub-eV scalar dark matter through the
  super-renormalizable Higgs portal}},
  \href{https://doi.org/10.1103/PhysRevD.82.043533}{\emph{Phys. Rev.}
  {\bfseries D82} (2010) 043533}
  [\href{https://arxiv.org/abs/1003.2313}{{\ttfamily 1003.2313}}].

\bibitem{Arvanitaki:2014faa}
A.~Arvanitaki, J.~Huang and K.~Van~Tilburg, \emph{{Searching for dilaton dark
  matter with atomic clocks}},
  \href{https://doi.org/10.1103/PhysRevD.91.015015}{\emph{Phys. Rev.}
  {\bfseries D91} (2015) 015015}
  [\href{https://arxiv.org/abs/1405.2925}{{\ttfamily 1405.2925}}].

\bibitem{Stadnik:2014tta}
Y.~V. Stadnik and V.~V. Flambaum, \emph{{Searching for dark matter and
  variation of fundamental constants with laser and maser interferometry}},
  \href{https://doi.org/10.1103/PhysRevLett.114.161301}{\emph{Phys. Rev. Lett.}
  {\bfseries 114} (2015) 161301}
  [\href{https://arxiv.org/abs/1412.7801}{{\ttfamily 1412.7801}}].

\bibitem{Arvanitaki:2015iga}
A.~Arvanitaki, S.~Dimopoulos and K.~Van~Tilburg, \emph{{Sound of Dark Matter:
  Searching for Light Scalars with Resonant-Mass Detectors}},
  \href{https://doi.org/10.1103/PhysRevLett.116.031102}{\emph{Phys. Rev. Lett.}
  {\bfseries 116} (2016) 031102}
  [\href{https://arxiv.org/abs/1508.01798}{{\ttfamily 1508.01798}}].

\bibitem{Graham:2015ifn}
P.~W. Graham, D.~E. Kaplan, J.~Mardon, S.~Rajendran and W.~A. Terrano,
  \emph{{Dark Matter Direct Detection with Accelerometers}},
  \href{https://doi.org/10.1103/PhysRevD.93.075029}{\emph{Phys. Rev.}
  {\bfseries D93} (2016) 075029}
  [\href{https://arxiv.org/abs/1512.06165}{{\ttfamily 1512.06165}}].

\bibitem{Geraci:2016fva}
A.~A. Geraci and A.~Derevianko, \emph{{Sensitivity of atom interferometry to
  ultralight scalar field dark matter}},
  \href{https://doi.org/10.1103/PhysRevLett.117.261301}{\emph{Phys. Rev. Lett.}
  {\bfseries 117} (2016) 261301}
  [\href{https://arxiv.org/abs/1605.04048}{{\ttfamily 1605.04048}}].

\bibitem{Arvanitaki:2016fyj}
A.~Arvanitaki, P.~W. Graham, J.~M. Hogan, S.~Rajendran and K.~Van~Tilburg,
  \emph{{Search for light scalar dark matter with atomic gravitational wave
  detectors}}, \href{https://doi.org/10.1103/PhysRevD.97.075020}{\emph{Phys.
  Rev.} {\bfseries D97} (2018) 075020}
  [\href{https://arxiv.org/abs/1606.04541}{{\ttfamily 1606.04541}}].

\bibitem{Arvanitaki:2017nhi}
A.~Arvanitaki, S.~Dimopoulos and K.~Van~Tilburg, \emph{{Resonant absorption of
  bosonic dark matter in molecules}},
  \href{https://arxiv.org/abs/1709.05354}{{\ttfamily 1709.05354}}.

\bibitem{Hees:2018fpg}
A.~Hees, O.~Minazzoli, E.~Savalle, Y.~V. Stadnik and P.~Wolf, \emph{{Violation
  of the equivalence principle from light scalar dark matter}},
  \href{https://doi.org/10.1103/PhysRevD.98.064051}{\emph{Phys. Rev.}
  {\bfseries D98} (2018) 064051}
  [\href{https://arxiv.org/abs/1807.04512}{{\ttfamily 1807.04512}}].

\bibitem{Geraci:2018fax}
A.~A. Geraci, C.~Bradley, D.~Gao, J.~Weinstein and A.~Derevianko,
  \emph{{Searching for ultra-light dark matter with optical cavities}},
  \href{https://arxiv.org/abs/1808.00540}{{\ttfamily 1808.00540}}.

\bibitem{Choi:2016luu}
K.~Choi and S.~H. Im, \emph{{Constraints on Relaxion Windows}},
  \href{https://doi.org/10.1007/JHEP12(2016)093}{\emph{JHEP} {\bfseries 12}
  (2016) 093} [\href{https://arxiv.org/abs/1610.00680}{{\ttfamily
  1610.00680}}].

\bibitem{Flacke:2016szy}
T.~Flacke, C.~Frugiuele, E.~Fuchs, R.~S. Gupta and G.~Perez,
  \emph{{Phenomenology of relaxion-Higgs mixing}},
  \href{https://doi.org/10.1007/JHEP06(2017)050}{\emph{JHEP} {\bfseries 06}
  (2017) 050} [\href{https://arxiv.org/abs/1610.02025}{{\ttfamily
  1610.02025}}].

\bibitem{Banerjee:2018xmn}
A.~Banerjee, H.~Kim and G.~Perez, \emph{{Coherent relaxion dark matter}},
  \href{https://arxiv.org/abs/1810.01889}{{\ttfamily 1810.01889}}.

\bibitem{Banerjee:2019epw}
A.~Banerjee, D.~Budker, J.~Eby, H.~Kim and G.~Perez, \emph{{Relaxion Stars and
  their detection via Atomic Physics}},
  \href{https://arxiv.org/abs/1902.08212}{{\ttfamily 1902.08212}}.

\bibitem{Linde:1986fd}
A.~D. Linde, \emph{{Eternally Existing Selfreproducing Chaotic Inflationary
  Universe}}, \href{https://doi.org/10.1016/0370-2693(86)90611-8}{\emph{Phys.
  Lett.} {\bfseries B175} (1986) 395}.

\bibitem{Goncharov:1987ir}
A.~S. Goncharov, A.~D. Linde and V.~F. Mukhanov, \emph{{The Global Structure of
  the Inflationary Universe}},
  \href{https://doi.org/10.1142/S0217751X87000211}{\emph{Int. J. Mod. Phys.}
  {\bfseries A2} (1987) 561}.

\bibitem{Nakao:1988yi}
K.-i. Nakao, Y.~Nambu and M.~Sasaki, \emph{{Stochastic Dynamics of New
  Inflation}}, \href{https://doi.org/10.1143/PTP.80.1041}{\emph{Prog. Theor.
  Phys.} {\bfseries 80} (1988) 1041}.

\bibitem{Nambu:1988je}
Y.~Nambu and M.~Sasaki, \emph{{Stochastic Approach to Chaotic Inflation and the
  Distribution of Universes}},
  \href{https://doi.org/10.1016/0370-2693(89)90385-7}{\emph{Phys. Lett.}
  {\bfseries B219} (1989) 240}.

\bibitem{Starobinsky:1994bd}
A.~A. Starobinsky and J.~Yokoyama, \emph{{Equilibrium state of a
  selfinteracting scalar field in the De Sitter background}},
  \href{https://doi.org/10.1103/PhysRevD.50.6357}{\emph{Phys. Rev.} {\bfseries
  D50} (1994) 6357} [\href{https://arxiv.org/abs/astro-ph/9407016}{{\ttfamily
  astro-ph/9407016}}].

\bibitem{Vanchurin:1999iv}
V.~Vanchurin, A.~Vilenkin and S.~Winitzki, \emph{{Predictability crisis in
  inflationary cosmology and its resolution}},
  \href{https://doi.org/10.1103/PhysRevD.61.083507}{\emph{Phys. Rev.}
  {\bfseries D61} (2000) 083507}
  [\href{https://arxiv.org/abs/gr-qc/9905097}{{\ttfamily gr-qc/9905097}}].

\bibitem{Creminelli:2008es}
P.~Creminelli, S.~Dubovsky, A.~Nicolis, L.~Senatore and M.~Zaldarriaga,
  \emph{{The Phase Transition to Slow-roll Eternal Inflation}},
  \href{https://doi.org/10.1088/1126-6708/2008/09/036}{\emph{JHEP} {\bfseries
  09} (2008) 036} [\href{https://arxiv.org/abs/0802.1067}{{\ttfamily
  0802.1067}}].

\bibitem{Dubovsky:2011uy}
S.~Dubovsky, L.~Senatore and G.~Villadoro, \emph{{Universality of the Volume
  Bound in Slow-Roll Eternal Inflation}},
  \href{https://doi.org/10.1007/JHEP05(2012)035}{\emph{JHEP} {\bfseries 05}
  (2012) 035} [\href{https://arxiv.org/abs/1111.1725}{{\ttfamily 1111.1725}}].

\bibitem{Graham:2018jyp}
P.~W. Graham and A.~Scherlis, \emph{{Stochastic axion scenario}},
  \href{https://doi.org/10.1103/PhysRevD.98.035017}{\emph{Phys. Rev.}
  {\bfseries D98} (2018) 035017}
  [\href{https://arxiv.org/abs/1805.07362}{{\ttfamily 1805.07362}}].

\bibitem{East:2016anr}
W.~E. East, J.~Kearney, B.~Shakya, H.~Yoo and K.~M. Zurek, \emph{{Spacetime
  Dynamics of a Higgs Vacuum Instability During Inflation}},
  \href{https://doi.org/10.1103/PhysRevD.95.023526}{\emph{Phys. Rev.}
  {\bfseries D95} (2017) 023526}
  [\href{https://arxiv.org/abs/1607.00381}{{\ttfamily 1607.00381}}].

\bibitem{Ooguri:2006in}
H.~Ooguri and C.~Vafa, \emph{{On the Geometry of the String Landscape and the
  Swampland}},
  \href{https://doi.org/10.1016/j.nuclphysb.2006.10.033}{\emph{Nucl. Phys.}
  {\bfseries B766} (2007) 21}
  [\href{https://arxiv.org/abs/hep-th/0605264}{{\ttfamily hep-th/0605264}}].

\bibitem{Blumenhagen:2018nts}
R.~Blumenhagen, D.~Kläwer, L.~Schlechter and F.~Wolf, \emph{{The Refined
  Swampland Distance Conjecture in Calabi-Yau Moduli Spaces}},
  \href{https://doi.org/10.1007/JHEP06(2018)052}{\emph{JHEP} {\bfseries 06}
  (2018) 052} [\href{https://arxiv.org/abs/1803.04989}{{\ttfamily
  1803.04989}}].

\bibitem{Ooguri:2018wrx}
H.~Ooguri, E.~Palti, G.~Shiu and C.~Vafa, \emph{{Distance and de Sitter
  Conjectures on the Swampland}},
  \href{https://arxiv.org/abs/1810.05506}{{\ttfamily 1810.05506}}.

\bibitem{ArkaniHamed:2006dz}
N.~Arkani-Hamed, L.~Motl, A.~Nicolis and C.~Vafa, \emph{{The String landscape,
  black holes and gravity as the weakest force}},
  \href{https://doi.org/10.1088/1126-6708/2007/06/060}{\emph{JHEP} {\bfseries
  06} (2007) 060} [\href{https://arxiv.org/abs/hep-th/0601001}{{\ttfamily
  hep-th/0601001}}].

\bibitem{Rudelius:2015xta}
T.~Rudelius, \emph{{Constraints on Axion Inflation from the Weak Gravity
  Conjecture}}, \href{https://doi.org/10.1088/1475-7516/2015/09/020,
  10.1088/1475-7516/2015/9/020}{\emph{JCAP} {\bfseries 1509} (2015) 020}
  [\href{https://arxiv.org/abs/1503.00795}{{\ttfamily 1503.00795}}].

\bibitem{Brown:2015iha}
J.~Brown, W.~Cottrell, G.~Shiu and P.~Soler, \emph{{Fencing in the Swampland:
  Quantum Gravity Constraints on Large Field Inflation}},
  \href{https://doi.org/10.1007/JHEP10(2015)023}{\emph{JHEP} {\bfseries 10}
  (2015) 023} [\href{https://arxiv.org/abs/1503.04783}{{\ttfamily
  1503.04783}}].

\bibitem{Heidenreich:2015wga}
B.~Heidenreich, M.~Reece and T.~Rudelius, \emph{{Weak Gravity Strongly
  Constrains Large-Field Axion Inflation}},
  \href{https://doi.org/10.1007/JHEP12(2015)108}{\emph{JHEP} {\bfseries 12}
  (2015) 108} [\href{https://arxiv.org/abs/1506.03447}{{\ttfamily
  1506.03447}}].

\bibitem{Saraswat:2016eaz}
P.~Saraswat, \emph{{Weak gravity conjecture and effective field theory}},
  \href{https://doi.org/10.1103/PhysRevD.95.025013}{\emph{Phys. Rev.}
  {\bfseries D95} (2017) 025013}
  [\href{https://arxiv.org/abs/1608.06951}{{\ttfamily 1608.06951}}].

\bibitem{Hawking:1981fz}
S.~W. Hawking and I.~G. Moss, \emph{{Supercooled Phase Transitions in the Very
  Early Universe}},
  \href{https://doi.org/10.1016/0370-2693(82)90946-7}{\emph{Phys. Lett.}
  {\bfseries 110B} (1982) 35}.

\bibitem{Agrawal:1997gf}
V.~Agrawal, S.~M. Barr, J.~F. Donoghue and D.~Seckel, \emph{{The Anthropic
  principle and the mass scale of the standard model}},
  \href{https://doi.org/10.1103/PhysRevD.57.5480}{\emph{Phys. Rev.} {\bfseries
  D57} (1998) 5480} [\href{https://arxiv.org/abs/hep-ph/9707380}{{\ttfamily
  hep-ph/9707380}}].

\bibitem{Hellerman:2005yi}
S.~Hellerman and J.~Walcher, \emph{{Dark matter and the anthropic principle}},
  \href{https://doi.org/10.1103/PhysRevD.72.123520}{\emph{Phys. Rev.}
  {\bfseries D72} (2005) 123520}
  [\href{https://arxiv.org/abs/hep-th/0508161}{{\ttfamily hep-th/0508161}}].

\bibitem{Tegmark:2005dy}
M.~Tegmark, A.~Aguirre, M.~Rees and F.~Wilczek, \emph{{Dimensionless constants,
  cosmology and other dark matters}},
  \href{https://doi.org/10.1103/PhysRevD.73.023505}{\emph{Phys. Rev.}
  {\bfseries D73} (2006) 023505}
  [\href{https://arxiv.org/abs/astro-ph/0511774}{{\ttfamily
  astro-ph/0511774}}].

\bibitem{Arvanitaki:2016xds}
A.~Arvanitaki, S.~Dimopoulos, V.~Gorbenko, J.~Huang and K.~Tilburg, \emph{{A
  small weak scale from a small cosmological constant}},
  \href{https://doi.org/10.1007/JHEP05(2017)071}{\emph{JHEP} {\bfseries 05}
  (2017) 071} [\href{https://arxiv.org/abs/1609.06320}{{\ttfamily
  1609.06320}}].

\bibitem{Weinberg:1987dv}
S.~Weinberg, \emph{{Anthropic Bound on the Cosmological Constant}},
  \href{https://doi.org/10.1103/PhysRevLett.59.2607}{\emph{Phys. Rev. Lett.}
  {\bfseries 59} (1987) 2607}.

\bibitem{Coleman:1988tj}
S.~R. Coleman, \emph{{Why There Is Nothing Rather Than Something: A Theory of
  the Cosmological Constant}},
  \href{https://doi.org/10.1016/0550-3213(88)90097-1}{\emph{Nucl. Phys.}
  {\bfseries B310} (1988) 643}.

\bibitem{Martin:1997ns}
S.~P. Martin, \emph{{A Supersymmetry primer}},
  \href{https://arxiv.org/abs/hep-ph/9709356}{{\ttfamily hep-ph/9709356}}.

\bibitem{Martin:2001vx}
S.~P. Martin, \emph{{Two loop effective potential for a general renormalizable
  theory and softly broken supersymmetry}},
  \href{https://doi.org/10.1103/PhysRevD.65.116003}{\emph{Phys. Rev.}
  {\bfseries D65} (2002) 116003}
  [\href{https://arxiv.org/abs/hep-ph/0111209}{{\ttfamily hep-ph/0111209}}].

\bibitem{Barbieri:2006bg}
R.~Barbieri, L.~J. Hall, Y.~Nomura and V.~S. Rychkov, \emph{{Supersymmetry
  without a Light Higgs Boson}},
  \href{https://doi.org/10.1103/PhysRevD.75.035007}{\emph{Phys. Rev.}
  {\bfseries D75} (2007) 035007}
  [\href{https://arxiv.org/abs/hep-ph/0607332}{{\ttfamily hep-ph/0607332}}].

\bibitem{Batra:2003nj}
P.~Batra, A.~Delgado, D.~E. Kaplan and T.~M.~P. Tait, \emph{{The Higgs mass
  bound in gauge extensions of the minimal supersymmetric standard model}},
  \href{https://doi.org/10.1088/1126-6708/2004/02/043}{\emph{JHEP} {\bfseries
  02} (2004) 043} [\href{https://arxiv.org/abs/hep-ph/0309149}{{\ttfamily
  hep-ph/0309149}}].

\bibitem{Cirelli:2007xd}
M.~Cirelli, A.~Strumia and M.~Tamburini, \emph{{Cosmology and Astrophysics of
  Minimal Dark Matter}},
  \href{https://doi.org/10.1016/j.nuclphysb.2007.07.023}{\emph{Nucl. Phys.}
  {\bfseries B787} (2007) 152}
  [\href{https://arxiv.org/abs/0706.4071}{{\ttfamily 0706.4071}}].

\bibitem{Sirunyan:2017cwe}
{\scshape CMS} collaboration, \emph{{Search for supersymmetry in multijet
  events with missing transverse momentum in proton-proton collisions at 13
  TeV}}, \href{https://doi.org/10.1103/PhysRevD.96.032003}{\emph{Phys. Rev.}
  {\bfseries D96} (2017) 032003}
  [\href{https://arxiv.org/abs/1704.07781}{{\ttfamily 1704.07781}}].

\bibitem{Aaboud:2017vwy}
{\scshape ATLAS} collaboration, \emph{{Search for squarks and gluinos in final
  states with jets and missing transverse momentum using 36 fb$^{-1}$ of
  $\sqrt{s}=13$ TeV pp collision data with the ATLAS detector}},
  \href{https://doi.org/10.1103/PhysRevD.97.112001}{\emph{Phys. Rev.}
  {\bfseries D97} (2018) 112001}
  [\href{https://arxiv.org/abs/1712.02332}{{\ttfamily 1712.02332}}].

\bibitem{Brust:2011tb}
C.~Brust, A.~Katz, S.~Lawrence and R.~Sundrum, \emph{{SUSY, the Third
  Generation and the LHC}},
  \href{https://doi.org/10.1007/JHEP03(2012)103}{\emph{JHEP} {\bfseries 03}
  (2012) 103} [\href{https://arxiv.org/abs/1110.6670}{{\ttfamily 1110.6670}}].

\bibitem{Papucci:2011wy}
M.~Papucci, J.~T. Ruderman and A.~Weiler, \emph{{Natural SUSY Endures}},
  \href{https://doi.org/10.1007/JHEP09(2012)035}{\emph{JHEP} {\bfseries 09}
  (2012) 035} [\href{https://arxiv.org/abs/1110.6926}{{\ttfamily 1110.6926}}].

\bibitem{Sirunyan:2017xse}
{\scshape CMS} collaboration, \emph{{Search for top squark pair production in
  pp collisions at $ \sqrt{s}=13 $ TeV using single lepton events}},
  \href{https://doi.org/10.1007/JHEP10(2017)019}{\emph{JHEP} {\bfseries 10}
  (2017) 019} [\href{https://arxiv.org/abs/1706.04402}{{\ttfamily
  1706.04402}}].

\bibitem{Aaboud:2017ayj}
{\scshape ATLAS} collaboration, \emph{{Search for a scalar partner of the top
  quark in the jets plus missing transverse momentum final state at
  $\sqrt{s}$=13 TeV with the ATLAS detector}},
  \href{https://doi.org/10.1007/JHEP12(2017)085}{\emph{JHEP} {\bfseries 12}
  (2017) 085} [\href{https://arxiv.org/abs/1709.04183}{{\ttfamily
  1709.04183}}].

\bibitem{Sirunyan:2017pjw}
{\scshape CMS} collaboration, \emph{{Search for supersymmetry in proton-proton
  collisions at 13 TeV using identified top quarks}},
  \href{https://doi.org/10.1103/PhysRevD.97.012007}{\emph{Phys. Rev.}
  {\bfseries D97} (2018) 012007}
  [\href{https://arxiv.org/abs/1710.11188}{{\ttfamily 1710.11188}}].

\bibitem{Aaboud:2017hrg}
{\scshape ATLAS} collaboration, \emph{{Search for supersymmetry in final states
  with missing transverse momentum and multiple $b$-jets in proton-proton
  collisions at $ \sqrt{s}=13 $ TeV with the ATLAS detector}},
  \href{https://doi.org/10.1007/JHEP06(2018)107}{\emph{JHEP} {\bfseries 06}
  (2018) 107} [\href{https://arxiv.org/abs/1711.01901}{{\ttfamily
  1711.01901}}].

\bibitem{StopyraThesis}
S.~Stopyra, \emph{Higgs dynamics during inflation},  Master's thesis, Imperial
  College London, 2014.

\end{thebibliography}\endgroup

\end{document}